\documentclass[a4paper,11pt]{article}
\usepackage{graphicx}
\usepackage{epstopdf}
\usepackage{subfigure}
\usepackage{dcolumn}
\usepackage{amsfonts,amsmath,amssymb}
\usepackage{hyperref}
\usepackage{float}

\usepackage{epsfig,multicol,bbm}

\newcommand\fverb{\setbox\pippobox=\hbox\bgroup\verb}
\newcommand\fverbit{\egroup\item[\fbox{\unhbox\pippobox}]}

\newbox\pippobox

\DeclareMathOperator{\sech}{sech}
\begin{document}
\title{\bf $Z_2$ Massive Axions, Domain Walls and Inflation}
\author{Shahrokh Assyyaee\thanks{Electronic address: s\_assyyaee@sbu.ac.ir}\;,\;\;\;{Nematollah Riazi\thanks{Electronic address: n\_riazi@sbu.ac.ir}}
\\
\small Department of Physics, Shahid Beheshti University, G.C., Evin, Tehran 19839,  Iran}
\maketitle
\begin{abstract}
We have analyzed a $U(1)$ model which is broken explicitly to a $Z_2$ model. The proposal results in generating two stable domain walls, in contrast with the more common $N_{DW}=1$ version which is prevalently used to explain axion invisibility for $U_{PQ}(1)$ model. We have been meticulous to take into account any possible relation with previous studies, even if they apparently belong to different lines of research. Then we have scrutinized the domain wall properties of the model, proposing a rigorous approximate solution which fully satisfies boundary conditions and the static virial theorem simultaneously. Invoking the mentioned approximation, we have been able to obtain an analytical insight about the effect of parameters  on the domain wall features, particularly on their surface energy density which is of great importance in cosmological studies when one tries to avoid domain wall energy domination problem. Next, we have mainly focused on the likely inflationary scenarios of the model, including saddle point inflation, again insisting on analytical discussions to be able to follow the role of parameters. We have tried to categorize any inflationary scenario into the known categories to take advantage of the previous detailed studies under the inflationary topic over the decades. We have concluded that any successful inflationary scenario requires large fields definition for the model. Calculations are mainly done analytically and numerical results are used as supportive material. 
\end{abstract}
\section{Introduction}
Recent rigorous observations have provided an unprecedented
accuracy that has to be taken into account in any cosmological
modeling \cite{1}-\cite{prcp}. Nowadays, we have enough
discriminating data to investigate the practicability of a
proposed inflationary scenario precisely \cite{oti,icp}. It is
widely believed that the recent Planck data favors the simplest
inflationary models consisting of a single field slow-roll
\cite{icp}. Although some inflationary models always remain in
the valid domain \cite{epp}, many of them have been excluded due
to incorrect predictions particularly in the density perturbation
spectral index on the CMB as well as the power of primordial
gravitational waves \cite{plc}. This decisive information is at
our disposal now, thanks to several experiments and decades of
rehearsing on the issue. Simultaneously, we are witnessing some
remarkable experiments in particle physics and quantum
field theory. No one can doubt that cosmology and quantum field
theory are tightly bound and any achievement in one of them must
be considered as a clue for the other. There are many attempts to
find a QFT motivation as a decisive sign of an acceptable
inflationary scenario \cite{ppm}-\cite{rcia} and conversely, the
capability of a QFT paradigm to include the inflation, is supposed as a
supportive sign for the paradigm \cite{Ghribi,cps}. On the
other hand, after endorsement of the Higgs boson existence, the
last predicted particle of the standard model \cite{mf,hda}, there
are more attention on the inflationary capability of symmetry
breaking scenario \cite{mf},\cite{Hai}-\cite{hpe}. The measured Higgs mass
in the LHC also raised another problem: the Higgs mass and the top
quark mass together increase the chance of being in a metastable
vacuum for the electroweak theory \cite{tqm}-\cite{vss} .
Topologically, a discrete vacuum means domain wall production if
the symmetry breaking is \textbf{perfect} \cite{TD}-\cite{cbds}. We have
known after Zeldovich's 1975 paper \cite{Zel} that domain walls
are drastically in contradiction with the observed cosmic mean
energy density unless the domain wall energy density is low enough. Such
low energy scales never provide appropriate outline for a
successful inflation although the CMB residual dipole anisotropy might be explained using them \cite{ipap}. The domain wall problem also appears when one tries to solve the strong CP problem by means of introducing a new axion field \cite{CPAX}. Indeed, to explain invisibility of axions due to their weak coupling to matter one could hypothesize more quark spices than the usual standard model quarks, or assume two Higgs doublets. The latter case is more appealing in quantum field theory since it offers the modest possible extension to the standard model. Moreover, there are yet unconfirmed reports of observing the footprint of Higgs-like particles in ATLAS and CMS \cite{HHR}, which put the multi-Higgs theories under the spotlight. Assuming a two doublet Higgs scenario, one inevitably encounters even number of domain walls separated by strings. The number of appearing domain walls are two times the number of the generators. If the energy scale of domain walls is high enough such that Domain wall production precedes the inflation, then one has an explanation for not observing such walls, like what happens in magnetic monopoles. Domain walls also could leave no significant
remnant in the later stages if they disappear soon enough. There
are some known mechanisms for destructing a domain wall which
could operate alone or in combination with each other \cite{cs}.
The most famous one is assuming a metastable domain wall which
automatically tends to ruin it \cite{cs,cdwp}. Of course, the decay
time could be very long, for example the decay time of electroweak
metastable vacuum, in the case of existence, is of the order of
the age of the universe \cite{vss}. Potentially, unstable domain walls are also among the best candidates for justification of baryogenesis. Amid the other options one
can mention destabilizing a domain wall by another defect
collision or embarking the symmetron mechanism \cite{symdom}. There is a very
interesting idea that mini black holes could trigger the electroweak vacuum
decay in a similar way.
On the other hand, one can generalize the natural inflation to include a dynamical modulus in addition to the proposed angular field dynamics. This generalization, respectively, promotes for example $U_{PQ}(1)$ to the double fields potential in which the $U(1)$ symmetry is broken to $Z_n$ discrete symmetry \cite{AxCos}. In this regard, it worth to have an exhaustive analysis of the potential with explicit $U(R)\longrightarrow Z_n$ broken symmetry, to know both the inflationary behavior and possible domain wall properties.   

Here, we try a double field potential \cite{rzi} with two discrete
vacua as a toy model of domain wall formation and inflation,
trying to avoid rendering numerical results before having an
analytical picture. By this, we always keep the track of the model
parameters employing appropriate approximations. The assumed
potential is very close to the original Higgs potential
\cite{Higgs,SSB} except that the continuous $U(1)$ symmetry now is
broken into a $Z_2$ symmetry with two discrete vacuua to produce
domain walls \cite{EU}. To get more familiar with the domain wall
properties which potentially could be produced by the proposed
model, we first calculate its energy scale for a simple spacial
configuration. Results show that entering more
parameters into the model and making it more sophisticated
provides us with more freedom to control the wall energy scale
without decreasing the energy scale of the potential at the origin
significantly, in contrast to the Zeldovich's proposal \cite{Zel}.
Next, we discuss the most important possible scenarios in
which the potential could accomplish inflation, starting with a
complete analytic review of the simple symmetry breaking case
according to the recent data. We stick to the most prevailing
method of diagnostic in which the slow-roll parameters play the basic
role in the analysis \cite{LL1,Dod}. We also overcome the
difficulties of dealing with a double field inflation
\cite{nci}-\cite{cmf} by treating the potential as an equivalent
single field potential. It soon becomes clear that almost all the
scenarios are compatible with the famous hill-top new inflationary
models \cite{HT}-\cite{nias}. Such models of course are categorized
among the super-Planckian models with no attainable motivation
from known physics, but like other new inflationary models some
particular characteristics make them noticeable: They predict very
small primordial gravitational waves
\cite{plc},\cite{wmc}-\cite{olive}, much less than what one may hope for detection in a conceivable future. The other point
about hill-top models is that there are some techniques to arrange
them to work in the supergravity scope \cite{sbi}-\cite{PS}.

The outline of the paper is as follows:
In the next section, we have a comprehensive review on the previous studies which provide motivations for our survey on descending the $U(1)$ symmetry to the $Z_2$. We focus on fundamental theories skipping applied physics and condensed matter. It is interesting that these different theories are related due to a common characteristic of producing $Z_2$ domain walls from an original $U(1)$ symmetry. Then in the third section, we analyze the domain wall characteristics of the proposed potential, where we find a very close approximation as a solution for the domain walls. This approximate solution satisfies the PDE's and static virial theorem simultaneously. Therefore, we invoke this approximation to deal with the other important domain wall characteristics, including the surface energy density and the wall thickness. Afterwards, in the forth section, we propose the potential as the source of inflation, starting from a novel analysis on the ordinary $U(1)$ symmetry breaking inflation. We try to have a complete survey on all possible scenarios of a successful inflation. We conclude that the saddle point inflation could reduce the scale of the required energy for inflation. This reduction being not such effective to avoid the theory from becoming a super-Planckian. The last section is devoted to the conclusion which contains the most important results of the paper.

\section{Motivations}
Although $Z_2$ symmetry could not assumed as a formal part of the standard model of particles, it appears frequently for certain reasons. One of these arenas is extending the Higgs sector. in fact, in spite of unnecessity of extra bosons in the standard model, many pioneering theories like Supersymmetry and grand unified theories, demand for extending the Higgs sector. The simplest and the most well-known extended models demand for two Higgs doublet models (2HDMs). 2HDMs also provide one of the best explanations for axion invisibility \cite{CPAX,DFSZ}. Axions are Nambo-Goldstone bosons of Pecci-Quinn $U(1)$ spontaneous symmetry breaking which originally was invented to solve the strong CP conservation problem \cite{CCPP}. Theoretically, spontaneous breaking of the $U_{PQ}(1)$ leaves domain wall(s) attaching to a string, \cite{AxCos}. Assuming a multi-Higgs model to deal with the recent unconfirmed record of observing a Higgs-like bump in LHC's last run \cite{HHR}, reinforces the existence of such extensions of the standard model Higgs sector. If the Higgs cousins contain interaction with standard model fermions, which is a very natural postulate, then one can assume $Z_2$ symmetry and the appropriate fermionic eigen value to avoid Higgs-mediated flavor changing neutral current (FCNC) \cite{THDM,LDMZ2}. Recently, a survey has proposed the cosmological consequences of explicit breaking of $U_{PQ}(1)$ \cite{AxCos}, considering the potential to be
\begin{equation}
V(\Phi)=\frac{\lambda}{4}\left({\Phi^\dagger}\Phi-\sigma^2\right)^2+\frac{m^2\sigma^2}{N_{DW}^2}(1-\cos{N_{DW}\theta}),
\end{equation}
where $\Phi$ is a complex field defined as $\Phi=\lvert\Phi\rvert e^{i\theta}$. In their analysis, the responsible term for breaking the $U(1)$ symmetry demonstrates only the phase field dependence. In our survey we let the explicit symmetry breaking term to have modulus dependence, too. We therefore assume
\begin{equation}
V(\Phi)=\frac{\lambda}{4}\left({\Phi^\dagger}\Phi-\sigma^2\right)^2+\frac{m^2\lvert\Phi\rvert^2}{N_{DW}^2}(1-\cos{N_{DW}\theta}).
\end{equation}
Also for simplicity, we focus on $N_{DW}=2$ case since it suffices to inspect the topological domain wall behavior. We are thus led to
\begin{equation}
V(\Phi)=\frac{\lambda}{4}\left({\Phi^\dagger}\Phi-\sigma^2\right)^2+\frac{m^2\lvert\Phi\rvert^2}{2}\sin ^2 \theta.
\label{ver1}
\end{equation}
It is worth mentioning that the above potential form is also a conformally renormalizable extension of natural inflation \cite{freese}. Natural inflation, was originally, based on the dynamics of the phase of a complex field whose modulus is stabilized severely. Then the Numbo-Goldstone boson becomes massive thanks to the instanton effect. In the QCD case, instantons break the $U(1)$ symmetry down to a discrete subgroup to produce the axion-inflaton potential
\begin{equation}
V(\phi)=\Lambda^4 (1+cos(\phi/f)).
\end{equation}
The key prediction of the above form of inflation is the strong gravitational wave remnant. In fact, strong enough to be detected in the recent Planck project. Lacking such approval from observation, one could suppose more completion to the original natural model. Bestowing modulus dependence on the potential is supposed to be one of the first choices. This choice has also been considered recently in \cite{TFRNI} by introducing 
\begin{equation}
V(\Phi)=\frac{\lambda}{4}\left(\Phi^\dagger\Phi-\sigma^2\right)^2+\Lambda^3(\Phi+\bar{\Phi}).
\end{equation}
The above potential coincides with KSVZ \cite{CPAX} modification of Pecci-Quinn theory, in which $N_{DW}=1$. In order to promote the above potential to contain two domain walls, which is more desirable in our study, we suggest
\begin{equation}
V(\Phi)=\frac{\lambda}{4}\left(\Phi^\dagger\Phi-\sigma^2\right)^2+\Lambda^2(\Phi+\bar{\Phi})^2.
\end{equation}
Then assuming $\Phi=\lvert\Phi\rvert e^{i(\pi/2-\theta)}$, one obtains
\begin{equation}
V(\Phi)=\frac{\lambda}{4}\left({\Phi^\dagger}\Phi-\sigma^2\right)^2+4\Lambda^2\lvert\Phi\rvert^2\sin ^2 \theta,
\label{ver2}
\end{equation}
which is just (\ref{ver1}), with a renaming of the parameters. As it will be introduced later in the paper, our choice of variables for dealing with (\ref{ver1}) or (\ref{ver2}) is
\begin{equation}
V(R,\Theta)=\frac{\lambda}{4}\left(R^2-\sigma^2\right)^2+\frac{\epsilon}{2}r^2 sin ^2 \Theta,
\label{ver3}
\end{equation}
where 
\begin{equation}
R^2\equiv\Phi^\dagger\Phi=(\phi-i\psi)(\phi+i\psi)=\phi^2+\psi^2
\end{equation}
and
\begin{equation}
\Theta\equiv\tan ^{-1} \frac{\psi}{\phi}.
\end{equation}
From a completely different point of view, in the inflationary paradigm of cosmology, there is a category of potentials, dubbed new inflation, in which the slow roll starts from nearly flat maximum of the potential where the field(s) is(are) located near the origin. In other words, the slow roll happens to be outward from the origin. These category of inflationary models, survived the tests though they generally suffer super-Planckian parameters. Among new inflationary theories one could mention inverted hybrid inflation \cite{MVHI}, which was an attempt to merge new inflation with the hybrid inflation. The potential has the following form
\begin{equation}
V(\phi,\psi)=V_0-\frac{1}{2}m_\phi ^2\phi^2+\frac{1}{2}m_\psi ^2\psi^2-\frac{1}{2}\lambda\phi^2\psi^2+\frac{1}{4}\lambda_\phi\phi ^4+\frac{1}{4}\lambda_\psi \psi ^4.
\label{inverted}
\end{equation} 
Now if one redefines the parameters as $\lambda=\lambda_\phi=\lambda_\psi$, $m_\phi ^2=\lambda\sigma^2$, $m_\psi ^2=\epsilon-\lambda\sigma^2$ and $V_0=\lambda\sigma^4 /4$ then (\ref{inverted}) reduces to
\begin{equation}
V(\phi,\psi)=\frac{\lambda}{4}\left(\phi^2+\psi^2-\phi_0^2\right)^2+\frac{\epsilon}{2}\psi^2.
\end{equation} 
which is just the Cartesian form of (\ref{ver3}). Of course in order to restore the tachyonic instability of the field $\psi$, an additional constraint $\epsilon>\lambda\sigma^2$ is needed, but in our study the latter condition won't be necessary. 
\section{ Supersymmetry and explicit breaking of the norm-space $U(1)$ symmetry}
In this sectin, we present a brief motivation for explicit $U(1)$ symmetry breaking from SUSY. To avoid lengthening the article, we avoid any introductory entrance to supersymmetry. The reader may consult many comprehensive textbooks on the subject. Supersymmetry, if exists at all, must be a broken symmetry. Many attempts have been made to introduce a viable explicit or spontaneous mechanism to explain the supersymmetry breaking. Here, we consider a D-term SSB by adding an additional $U(1)$ gauge symmetry (Fayet-Iliopoulos mechanism) \cite{FIM} and derive the resultant potential. We will see that the resulting potential, for the case of two charged scalar fields, demonstrates an explicitly broken $U(1)$ behavior, when is exhibited in norm-space of the fields. Moreover, since LHC has been obtained no approval evidence for minimal supersymmetric standard model (MSSM) \cite{LHCB}, considering additional fields to the supersymmetry sounds as a next logical step. One of the elegant properties of supersymmetry is the automatic appearance of a scalar potential through F and D auxiliary fields which are originally invented to balance the off-shell bosonic and fermionic degrees of freedom \cite{WBP,SSPM}: 
\begin{equation}
V=V_D+V_F=\frac{1}{2}\sum_a \lvert D^a\rvert^2+\sum_i \lvert F_i\rvert^2
\end{equation} 
Supersymmetry requires the vacuume expectation value (vev) to vanish. In this regard, non-vanishing (positive) vev is considered as a sign for SSB. In other words, if both D-term and F-term super potential contributions can't be zero coincidently, then the supersymmetry is broken. One way to do this task is assuming an extra $U(1)$ symmetry and let the potential to involve a linear D-term besides the ordinary terms:
\begin{equation}
V=\kappa D-\frac{1}{2}D^2-g D\sum_j q_j\lvert A_j\rvert^2,
\end{equation} 
where $A_j$ represents the fields that acquire charge under new $U(1)$ symmetry. The first term, known as Fayet-Iliopoulos (FI) term, satisfies both supersymmetry and $U(1)$ gauge symmetry. Then supposing the charged scalar field to be massive, one obtains
\begin{equation}
V=\sum_j \lvert m_j\rvert^2\lvert A_j\rvert^2+\frac{1}{2}\left(\kappa-g\sum_j q_j\lvert A_j\rvert^2\right)^2.
\label{FIL}
\end{equation}
Obviously, the above potential has a non-zero minimum which demonstrates breaking of supersymmetry. One has to note that despite broken supersymmetry, the $U(1)$ gauge remains unbroken if $\lvert m_j\rvert^2>\kappa gq_j$ for all fields. If $A_j$ for all $j$'s happen to be massive, then they must appear in pairs with apposite charges to respect the gauge symmetry. Of course, there is no obligation for charged scalar fields to be massive since they can be considered massless without losing any bosonic degrees of freedom. In order to get closer to the model considered in this paper, we consider two massive charged scalar fields $\phi$ and $\psi$ and recast $\lambda=2g^2$ and $\phi_0 ^2=k/g$ while the charges are normalized to $\pm 1$. We obtain
\begin{equation}
V=\frac{\lambda}{4}\left(\lvert\phi\rvert^2-\lvert\psi\rvert^2-\phi_0 ^2\right)^2+m_\phi ^2 \lvert\phi\rvert^2+m_\psi ^2 \lvert\psi\rvert^2.
\label{DSSB}
\end{equation}
Let us redefine the potential in polar coordinates by setting $\lvert\phi\rvert= R\cos\Theta$ and $\lvert\psi\rvert= R\sin\Theta$, then we have
\begin{equation}
V=\frac{\lambda}{4}\left(R^2 \cos 2\Theta-\phi_0 ^2\right)^2+m_\phi ^2 R^2 \cos^2\Theta+m_\psi ^2 R^2 \sin^2\Theta.
\end{equation}
Taking derivative with respect to angular coordinate yields
\begin{equation}
\frac{\partial V}{\partial \Theta}=R^2 \sin{2\Theta}\left[\lambda\left(\phi_0 ^2-R^2\cos{2\Theta}\right)+m_\psi ^2-m_\phi ^2\right]
\end{equation}
Then two groups of solutions will be gained for $\frac{\partial V}{\partial \Theta}=0$. $\Theta=k\pi/2$ and  $\Theta=\frac{1}{2}\cos^{-1}\left(\frac{m_\psi ^2-m_\phi ^2}{\lambda R^2}+\frac{\phi_0 ^2}{R^2}\right)$. The former is always available but the latter needs an elaborated fine tuning of parameters, particularly, when the $R$ field is located near the origin initially, like what happens in a normal new inflationary scenario. Therefore, one can consider four attractor paths down from the origin as the slow-roll path, which are equal pairwise.
\begin{equation}
\Theta=(2k+1)\frac{\pi}{2}\Longrightarrow V=\frac{\lambda}{4}\left(R^2 -\phi_0 ^2\right)^2+m_\psi ^2 R^2=\frac{\lambda}{4}\left[R^2-(\phi_0 ^2-\frac{2}{\lambda}m_\psi ^2) \right]^2+m_\psi ^2 (\phi_0 ^2-\frac{m_\psi ^2}{\lambda}),
\end{equation}    
\begin{equation}
\Theta=(2k)\frac{\pi}{2}\Longrightarrow V=\frac{\lambda}{4}\left(R^2 -\phi_0 ^2\right)^2+m_\phi ^2 R^2=\frac{\lambda}{4}\left[R^2-(\phi_0 ^2-\frac{2}{\lambda}m_\phi ^2) \right]^2+m_\phi ^2 (\phi_0 ^2-\frac{m_\phi ^2}{\lambda}).
\end{equation}   
To protect the $U(1)$ gauge symmetry for each field one requires $\phi_0<\sqrt{2/\lambda}m_j ^2$, where $j$ stands for the corresponding field. One has to note that the potential exhibits the SSB due to its positive definition, whether the gauge symmetry has been broken or not. In the case of $U(1)$ gauge symmetry breaking, the potential develops one-dimensional kinks, different in shape for each scalar. If both scalar fields are involved in the gauge symmetry breaking, then four vacua will develop. For $m_\psi=m_\phi$, all vacua are degenerate and stable, while for $m_\psi\neq m_\phi$, the two vacua which are related to the more massive field $vev_{(meta stable)}=m_{(>)} (\phi_0 ^2-\frac{m_{(>)}}{\lambda})$, become metastable and ultimately decay to the two stable vacua $vev_{(stable)}=m_{(<)} (\phi_0 ^2-\frac{m_{(<)}}{\lambda})$. Since the decay rate decreases exponentially by decreasing the difference of vev's, it is possible to consider the situations in which the metastable lifetime exceeds the age of the universe.  
Here, some more elaboration is in order. If (\ref{DSSB}) is assumed for supersymmetric masses set to zero then one obtains
\begin{equation}
V=\frac{\lambda}{4}\left(\lvert\phi\rvert^2-\lvert\psi\rvert^2-\phi_0 ^2\right)^2=\frac{\lambda}{4}\left(\lvert\phi\rvert^2+\lvert\psi\rvert^2-\phi_0 ^2\right)^2+\lambda\left(\phi_0^2 - \lvert\phi\rvert^2\right) \lvert\psi\rvert^2,
\end{equation}
which means that in the space of the charged scalar field norms, the potential shows a $U(1)$ symmetry, explicitly broken by an effective mass of one of the fields.
\section{A Two Fields Potential: Domain Wall Analysis }
\begin{figure}\hspace{4.cm}
\centering
\includegraphics[width=1\linewidth]{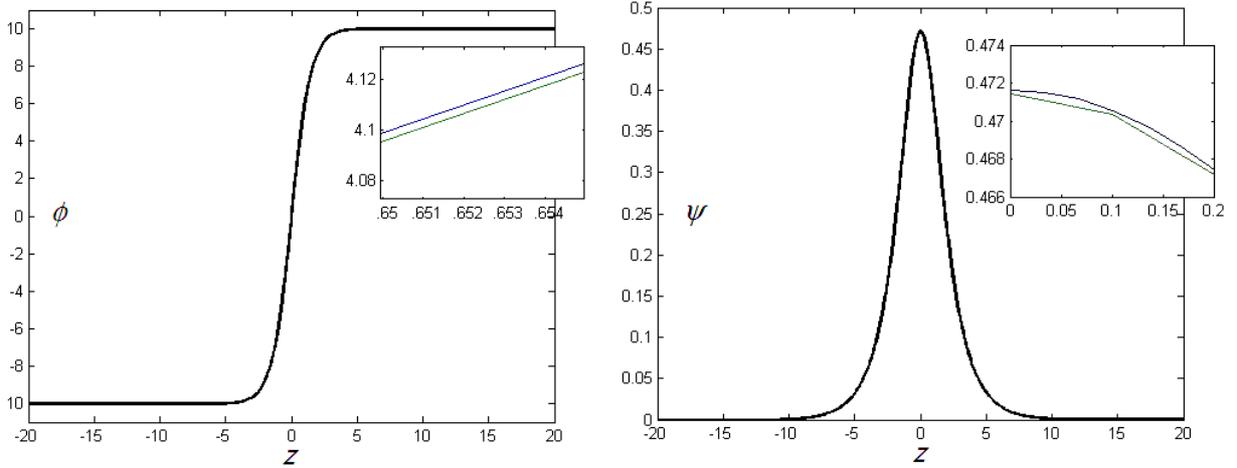}
\caption{In these figures $\phi(z)$ and $\psi(z)$ have been
plotted for $\lambda=0.009$, $\epsilon=0.449$ and $\phi_0=10$,
although the parameters ave been selected very close to the
boundary of the approximation validity ($\phi_0
^2=\frac{2\epsilon}{\lambda}$) still the analytical solutions are very close to numerical solutions. The insets are zoomed to demonstrate the tiny difference. } \label{fig:domainwall}
\end{figure}

To begin, we propose the simplest asymmetric scenario in a
two-fields potential in which we require that $V(\Phi,\Psi)$
satisfies the following constraint:
\begin{equation}
\frac{\partial V(\Phi,\Psi)}{ \partial \Psi} -\frac{\partial
V(\Phi,\Psi)}{\partial \Phi}=\epsilon.
\label{simplest}
\end{equation}
For $\epsilon=0$, we know that $V(\Phi,\Psi)=f(\phi+\psi)$ is a
solution of (\ref{simplest}) for any arbitrary function $f(x)$.
This ensures $U(1)$ symmetry if we choose $\Phi=\frac{\phi^2}{2}$
and $\Psi=\frac{\psi^2}{2}$, and consequently $\epsilon\neq 0$
breaks this symmetry. The above equation has the general solution
of the form
\begin{equation}
V(\Phi,\Psi)=f(\Phi+\Psi)+\epsilon \Psi,\label{wwe}
\end{equation}
or
\begin{equation}
V(\Phi,\Psi)=f(\Phi+\Psi)-\epsilon \Phi,
\end{equation}
where we consider positive definition for $\epsilon$ throughout the paper. To achieve a more familiar potential form let us recast the fields
into $\Phi=\frac{\phi^2}{2}$ and $\Psi=\frac{\psi^2}{2}$ and also
let the function $f(\phi,\psi)$ to have an ordinary symmetry
breaking appearance. Then the potential (\ref{wwe}) could be
written as
\begin{equation}
V(\phi,\psi)=\frac{\lambda}{4}\left(\phi^2+\psi^2-\phi_0^2\right)^2\pm\frac{\epsilon}{2}\psi^2.
\label{original}
\end{equation}
The potential (\ref{original}) shows a full circular symmetry in
the first parenthesis resembling the Higgs potential
\cite{Higgs,Ken}. In the inflationary context this is a self
consistent version of a particular hill-top model \cite{six}. In
fact, without the last term in (\ref{original}), the circular
freedom in the vacuum corresponds to the massless Goldstone boson
\cite{SSB} but for the case under consideration, the vacuum is not
a continuous minimum and as we will see, the circular field
acquires mass as well as its own roll down mechanism which has to
be considered in slow roll assumption and could bring about
different consequences. From now on, we will discuss the symmetry breaking term with the plus sign in (\ref{original}) , but a brief argument about this choice is in order. Suppose we choose the plus sign in (\ref{original}),
Then by the following redefinition
\begin{equation}
\phi_0 ^\prime\equiv\left(\frac{\epsilon}{\lambda}-\phi_0^2\right)^{1/2}
\end{equation}
and interchanging the variables $(\phi\rightleftharpoons\psi)$, one obtains
\begin{equation}
V(\phi,\psi)=V_0+\frac{\lambda}{4}\left(\phi^2+\psi^2-{\phi_0 ^\prime} ^2\right)^2-\frac{\epsilon}{2}\psi^2,
\end{equation} 
where
\begin{equation}
V_0\equiv\frac{\epsilon}{4}\left(2\phi_0 ^2-\frac{\epsilon}{\lambda}\right).
\end{equation}
So the plus or minus choice for epsilon coefficient is trivial up to a constant. 
Although breaking the $U(1)$ symmetry down
to $Z_2$ is very common in condense matter and superconductivity \cite{mtsc}, it has received less attention in
fundamental theories. In order to have a better perspective about
the potential, let us change the field coordinates into polar
coordinates by setting $\varphi=R\cos \Theta$ and $\psi=R\sin
\Theta$. The potential then becomes
\begin{equation}
 V(R,\Theta)=\frac{\lambda}{4}(R^2-\phi_0^2)^2+\frac{\epsilon}{2}R^2\sin^2\Theta.
\label{polar}
\end{equation}
Now we are able to discuss the behavior of the potential by taking
a differentiation with respect to the radial field.
\begin{equation}
\frac{\partial V(R,\Theta)}{\partial R}=\lambda
R(R^2-\varphi_0^2)+\epsilon R sin\Theta.
\end{equation}
To learn about the extrema, let us find the roots of
$\frac{\partial V}{\partial R}=0$ ;
\begin{equation}
R_{ext}=\{0,\pm\sqrt{\frac{\lambda\phi_0^2-\epsilon\sin^2\Theta}{\lambda}}\}
\end{equation}
Obviously, there could be up to three roots. Here we are able to
categorize the potential as bellow
\begin{equation}
\lambda\phi_0 ^2 >\epsilon \quad\xrightarrow{\rm 3 \;roots\; for
\;all\; \Theta 's}{\rm \quad local\; maximum\; at\; the\; origin}
\end{equation}
\begin{equation}
\lambda\phi_0 ^2 \leqslant\epsilon \quad\xrightarrow[\rm
\;\;depending \; on\; \Theta \;\;] {\rm 1 \; or\; 3 \; roots\;}
{\rm \quad saddle\; point\; at\; the\; origin}\label{ue2}
\end{equation}
\begin{center}
\begin{figure}\hspace{0.5cm}
 \includegraphics[width=0.45\linewidth]{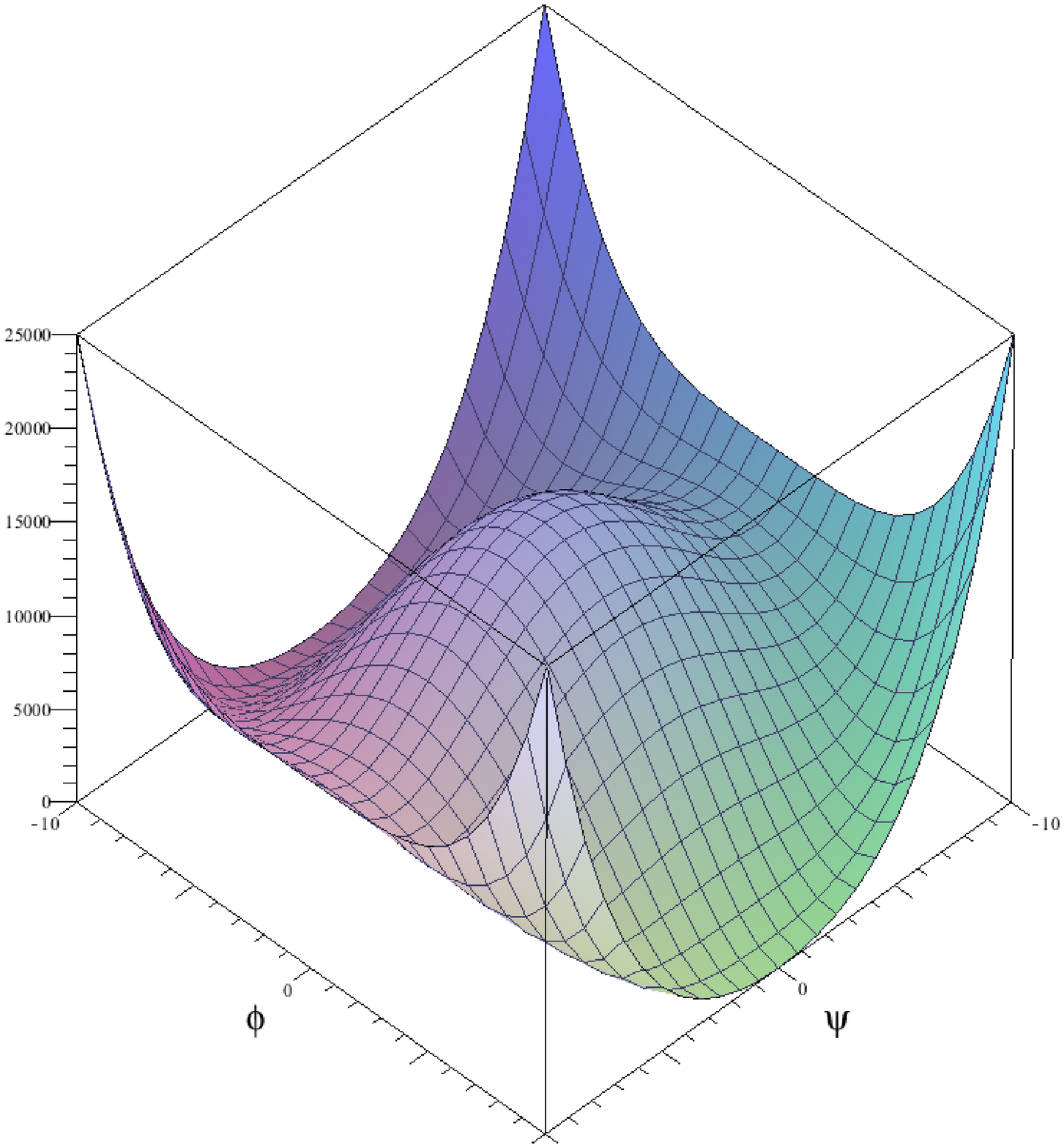}
 \includegraphics[width=0.45\linewidth]{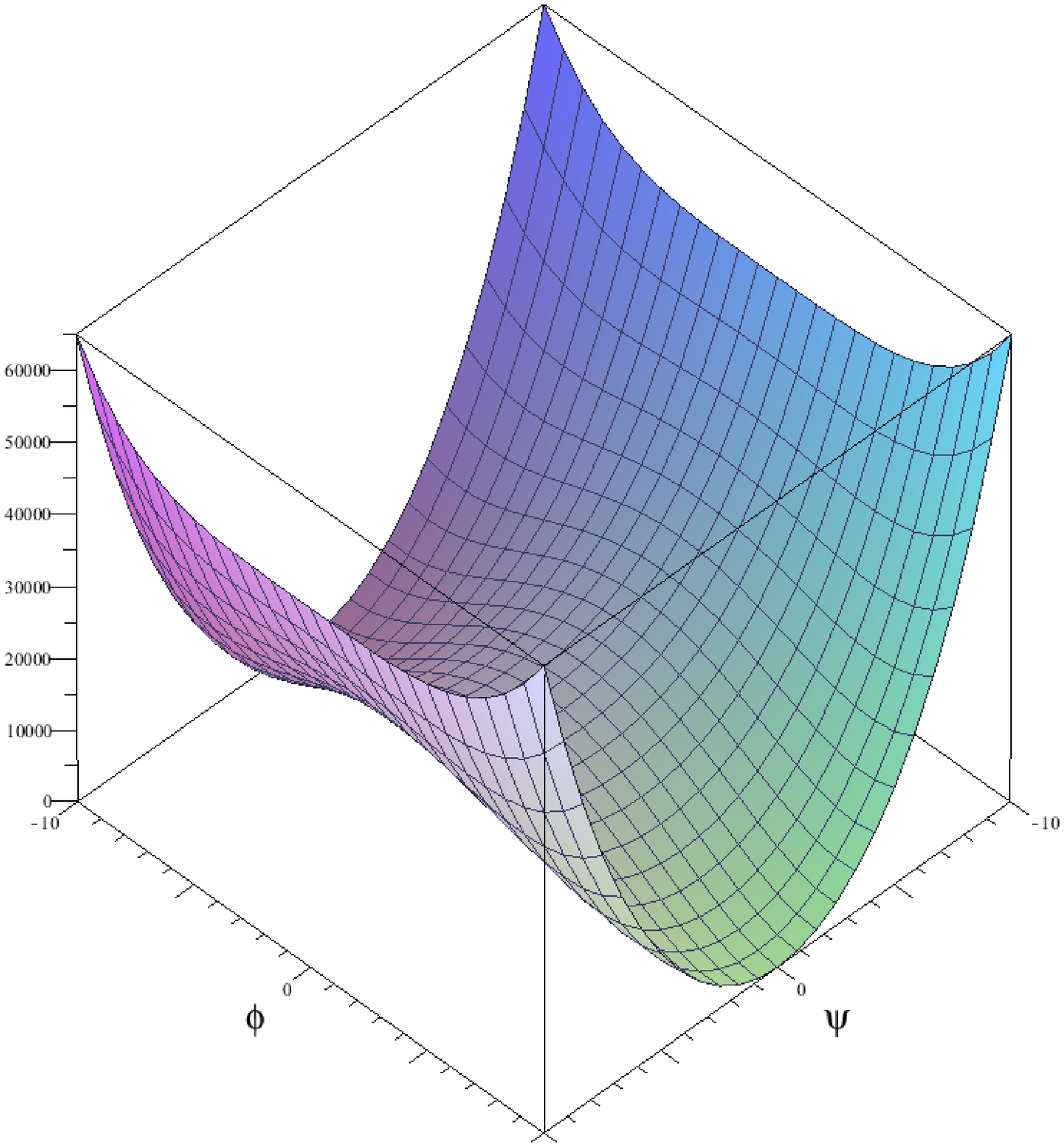}
\caption{These height plots indicate the potential shape for
$\lambda\phi_0 ^2
>\epsilon$ (left) and $\lambda\phi_0 ^2
<\epsilon$ (right). The origin appears as local maximum for
$\lambda\phi_0 ^2
>\epsilon$ and is a saddle point for the other case.  } \label{fig:lmsp}
\end{figure}
\end{center}
The stronger the inequality (\ref{ue2}), the wider range of
$\Theta$ recognizes the origin as minimum. For $\lambda\phi_0 ^2
>\epsilon$, the two saddle points are located at $\psi=\pm\sqrt{\phi_0 ^2-
\frac{\epsilon}{\lambda}}$ on the $\psi$ axis. As the inequality
$\lambda\phi_0 ^2>\epsilon$ becomes weaker the saddle points move
toward the origin where finally meet the origin for $\lambda\phi_0
^2=\epsilon$. The origin remains a saddle point for $\lambda\phi_0
^2 \leqslant\epsilon$ (Figure  \ref{fig:lmsp}). Since the potential
exhibits discrete vacua after symmetry breaking, domain wall
production seems inevitable. More technically, it is the vacuum
manifold $\mathcal{M}$ that determines the character of possible
topological defects \cite{TD} and in our case the zero homotopy
group, or zero homotopy set as mathematicians prefer, is not
trivial $(\pi_0(\mathcal{M})\neq 1)$ which warns us about the inevitability of domain walls generation through the Kibble mechanism
\cite{Kibble,Kibble2}. One expects at least one domain wall per
horizon volume if the symmetry breaking is {\bf perfect}. The
potential (\ref{original}) has recently been analyzed due to the
capability of generating domain walls with a rich dynamics
\cite{rzi}. There is a tight constraint on the existence of domain
walls except for very low surface energy densities \cite{Zel}. Let
us see why this is so. For the very popular toy model of $Z_2$
discrete symmetry breaking in which the effective potential has
the $\phi^4$ well-known form;
\begin{equation}
V(\phi)=\frac{\lambda}{4}\left(\phi^2-\phi_0^2\right)^2.
\label{Z2}
\end{equation}
where $\phi$ is a real field, the scenario has been already
analyzed in many references \cite{TD,kdw,cs,cstd} so it would be
adequate to review only the results. For simplicity, we assume a
Minkowskian background space-time since it suffices to indicate
the major properties. After the symmetry breaking is settled down, as
a very simple simulation, one could suppose a planar domain wall
placed on the $xy$ plane at $z=0$. For planar domain walls,
everything is independent of $x$ and $y$ coordinates and as long
as we are interested in the static situation, the time dependency
is eliminated. So we get
\begin{equation}
\partial^z\partial_z \phi - \frac{\partial V}{\partial
\phi}=0. \label{zuz}
\end{equation}
The first integral of this equation is
\begin{equation}
\frac{1}{2}\sum_i\,\,(\partial_z\phi)^2-V=C.
\end{equation}
where the constant of integration $C$ vanishes when we impose the
boundary conditions of vanishing of the potential and the spacial field
derivative at the infinity. Then the domain wall solution for the
assumed boundary condition is
\begin{equation}
z=\int \frac{d\phi}{\sqrt{2V(\phi)}}\; ,
\end{equation}
or
\begin{equation}
\phi(z)=\phi_0\,\tanh(\sqrt{\lambda / 2 }\, \phi_0 z).
\end{equation}
and for the surface energy density of the domain wall we obtain
\begin{equation}
\sigma=\int T^0\,_0 dz=\frac{4}{3}\lambda^{1/2}{\phi_0}^3.
\label{SD Toy}
\end{equation}
The thickness of the wall is defined as $ (\lambda / 2 )^{-1/2}\,
\phi_0^{-1}$. In an expanding universe, any proper velocity of the
walls very soon becomes negligible, which leaves the universe with a
non-relativistic network of domain walls, and here the problem
arises; According to the Kibble mechanism, domain walls are
generally horizon-sized so we can estimate their mass as if we do
for a horizon-sized plate i.e. $m_{wall}\sim\lambda^{1/2}\phi_0^3
H^{-2}$, so the mass energy density can roughly be approximated as
$\rho_{wall}\sim\phi_0 ^3 t^{-1}$. We know that the critical
density evolves as $\rho_{crit}\sim t^{-2}$. Therefore
$R_{wall}(t)\sim(\phi_0/M_{Pl})^2 \phi_0 t$. By setting $\phi_0$
at about the GUT scale, the domain wall energy density reaches the
critical density already in the time of wall generation and one
expects that in our time this ratio becomes $R_{wall}(t\sim
10^{18})\sim 10^{52}$ which shows a catastrophic conflict with
reality. In order to compromise between the introduced domain
walls and the observations, one needs to decrease the energy of
the possible domain walls to very small values \cite{Zel}. As we
will derive in (\ref{sedw}) for the more sophisticated domain
walls other parameters involve to determine the wall energy which
gives us sort of freedom to prepare the potential to work as
inflation. The other remedy is to allow the disappearance of
domain walls so early that not only diminish from the density
calculations but also not altering the CMB isotropy, considerably.
This can be achieved in various ways. For example, one can imagine
the potential as an effective potential to demote the wall to an
unstable version, by which, one of the vacua will disappear through
the biased tunneling effect, or considering some sort of
destructing collisions which are generally fatal for a kink
stability and for such primordial walls the primordial black holes
might be the best candidates \cite{freese}. It is worth mentioning that domain walls even at very low energies could cause a residual
dipole anisotropy in large scale observations, and such an
anisotropy is receiving increasing observational supports \cite{ipap}.
Static double-field domain wall solutions corresponding to
(\ref{original}) satisfy the following two coupled Euler-Lagrange
equations
\begin{equation}
\frac{\partial^2 \phi}{\partial
z^2}=\lambda\phi(\phi^2+\psi^2-\phi_0 ^2), \label{D1}
\end{equation}
\begin{equation}
\frac{\partial^2 \psi}{\partial
z^2}=\lambda\psi(\phi^2+\psi^2-\phi_0 ^2)+\epsilon\psi.\label{D2}
\end{equation}
These equations can be merged into
\begin{equation}
\frac{1}{\psi}\frac{\partial^2 \psi}{\partial
z^2}-\frac{1}{\phi}\frac{\partial^2 \phi}{\partial z^2}=\epsilon.
\label{merged}
\end{equation}
To find the solution of the above equation with appropriate
boundary conditions, we have a guide line; $\phi(z)$ must be odd
with respect to $z$ coordinate due to its main role in the
discrete symmetry breaking process, while for $\psi(z)$ both sides
of the wall in $z=0$ has the same characteristics since both vacua
lay at $\psi=0$. In other words, $\psi$ has to be even with
respect to the z coordinate. Moreover (\ref{merged}) is subject to the
following boundary conditions
\begin{equation}
\phi(z=0)=0,
\end{equation}
\begin{equation}
\phi(z=\pm\infty)=\pm \phi_0,
\end{equation}
\begin{equation}
\psi(z=\pm\infty)=0.
\end{equation}
The trajectory of transition between the vacua doesn't pass
through $(\phi=0, \psi=0)$ if $\phi_0^2>\epsilon/\lambda$, since
the origin in this case is a local maximum of the potential. To
find the domain wall solution, first, we employ an appropriate
ansatz for one field and derive the other field. Then we will check
the accuracy of the final solution by comparing it with the
numerical solutions. Our estimation about the final form must
fulfill the boundary conditions. The best choice would be
$\phi(z)=\phi_0\tanh{\sqrt{\epsilon}z}$, this hyperbolic form
which has been inspired by the $\phi^4$ kink solution, fully
satisfies the boundary conditions and indicates the odd
characteristic of $\phi(z)$. The appearance of $\sqrt{\epsilon}$
is also reasonable since after two times differentiation it will
produce the desired $\epsilon$ factor while the $\phi_0$
coefficient cancels out by division. Next, we put this ansatz
solution into (\ref{merged}), which after some straightforward
calculation one obtains for $\psi(z)$;
\begin{equation}
\psi(z)={C_1}{\sech{\sqrt{\epsilon}z}}+C_2{\sech{\sqrt{\epsilon}z}}({\sinh{2\sqrt{\epsilon}z}+2\sqrt{\epsilon}z}),\label{mge}
\end{equation}
where $C_1$ and $C_2$ are constants of integration. But the above
solution has been separated into an odd term and an even term. So
to keep the evenness property of $\psi(z)$ we require $C_2=0$. To
fix the solution we have to find the remaining constant $C_2$.
This can be done by means of the minimum energy theorem and
integration, but we utilize the static virial theorem, since both
of these two theorems stem from the least action principle, they
could be used interchangeably. The static virial theorem has
another important consequence of vanishing the tangential pressure
for the wall, which we prove before inserting it into our
calculation. For a typical multi-field potential the
Euler-Lagrange equations have the general form
\begin{equation}
\square\chi_a-\frac{\partial V}{\partial \chi_a}=0,
\end{equation}
where $a=\lbrace 1,2,... \rbrace$ enumerates the fields. For a
static solution and a planar wall these can be written as
\begin{equation}
\frac{d^2 \chi_a}{{dx_{_{\perp}}}^2}-\frac{\partial V}{\partial
\chi_a}=0,
\end{equation}
Here $x_{_{\perp}}\rightarrow z$ is chosen to be the coordinate
perpendicular to the wall. Multiplying by $\frac{d \chi^a}{dz}$
one obtains
\begin{equation}
\frac{d \chi^a}{dz}\left[\frac{d^2 \chi_a}{dz^2}-\frac{\partial
V}{\partial
\chi^a}\right]=\frac{d}{dz}\left[\frac{1}{2}\left(\frac{d\chi_a}{dz}\right)^2-V(\chi_a)\right]=0,
\end{equation}
\begin{equation}
\frac{1}{2}\sum_a\left(\frac{d\chi_a }{dz}\right)^2-V=constant.
\end{equation}
If we require a true vacuum to have zero energy expectation value
then the constant of integration in the above equation should be
zero. Note that the derived static one dimensional version of
virial theorem must be valid for the proposed guess if we require
it to satisfy the Euler-Lagrange equation of motion. Obtaining
another relation among the fields first derivative and the
potential, one can use it in order to determine $C_1$ in
(\ref{mge}) by requiring
\begin{equation}
\left(\frac{d \phi(z)}{d z}\right)^2+\left(\frac{d \psi(z)}{d
z}\right)^2=2V\left(\phi(z),\psi(z)\right).
\end{equation}
Substituting for $\phi(z)$, $\psi(z)$ and
$V\left(\phi(z),\psi(z)\right)$ and after some algebraic
simplification one finds $C_1=\sqrt{\phi_0 ^2-2\epsilon/\lambda}$
which results in
\begin{equation}
\phi(z)=\phi_0\tanh{\sqrt{\epsilon}z},\label{pdw}
\end{equation}
\begin{equation}
\psi(z)=\sqrt{\phi_0^2-2\frac{\epsilon}{\lambda}}\sech{\sqrt{\epsilon}z}.\label{sdw}
\end{equation}
This solution approximates very closely the more accurate
numerical solution (Figure  \ref{fig:domainwall}) and the wall width
is $\epsilon^{-1/2}$. As the next step one
can calculate the energy-momentum of the wall
\begin{equation}
T^\mu\;_\nu=\left[\left(\frac{\partial\phi}{\partial
z}\right)^2+\left(\frac{\partial\psi}{\partial
z}\right)^2\right]diag(1,1,1,0)=\frac{\epsilon\phi_0
^2}{\cosh^2{\sqrt{\epsilon}z}}\left(1-\frac{2\epsilon}{\lambda\sigma^2}\tanh^2{\sqrt{\epsilon}z}\right),
\end{equation}
Note that although our domain wall solution (\ref{pdw},\ref{sdw})
was approximate, the vanishing of $T^z\,_z=0$ is an exact
result \cite{Peeble}. If we use this result to find the
surface-energy density we find
\begin{equation}
\sigma\equiv\int_{-\infty} ^{+\infty}{T^0\,_0\, dz}=\int_{-\infty}
^{+\infty}{(\partial_z \phi)^2+(\partial_z
\psi)^2}\,dz=2\sqrt{\epsilon}\phi_0
^2\left(1-\frac{\epsilon}{3\lambda\phi_0 ^2}\right).\label{sedw}
\end{equation}
This result seems interesting since now one can control the effect
of $\phi_0$ by $\epsilon$ and ultimately for the case of
$\epsilon\ll\lambda\phi_0 ^2$ one obtains $\sigma\approx
2\sqrt{\epsilon}\phi_0 ^2$, which is independent of $\lambda$.
This result is noticeable because in contrary to the $\phi^4$
kink case, now it is possible to decrease the wall energy without
decreasing the maximum of the potential, In other words, the peak
could be chosen, say at the GUT or even Planck scale , since the wall energy
density is low enough to avoid domain wall density domination. To
be more clear parameter $\lambda$ controls the height of the
potential through $\lambda\phi_0/4$ while $\epsilon$ is
responsible for the wall energy density.

The above approximate solution, circumspectly, could be generalized to the below potentail
\begin{equation}
 V(R,\Theta)=\frac{\lambda}{4}(R^2-\phi_0^2)^2+\frac{\epsilon}{2}R^2\sin^2 \xi\Theta.
\label{gen}
\end{equation}
where $\xi$ is an arbitrary real number. In order to adapt our postulate to the well-analyzed $Z_n$ groups, we require $\xi$ to be integer or half integer. Since other choices of $\xi$ are related to our line through a field redefinition, the above assumption doesn't demote the level of the generality. Then one has to note that (\ref{gen}) shows explicit breaking of $U(1)$ symmetry into the $Z_{2\xi}$ where $\xi=1$ restores (\ref{polar}). By this, one obtains $2\xi$ domain walls, which are equal up to the free Lagrangian. But the picture becomes more contrived if one assumes interactions of $\phi(R,\Theta)$ and $\psi(R,\Theta)$ independently, since they acquire different values for each vacuum. Now, let us practice the (\ref{pdw},\ref{sdw}) approximation for the new case. Suppose we have a potential with $\xi=n_0/2$ which provides $n_0$ domain walls. first we recast the fields into their Cartesian form by $\phi\equiv R\cos n_0 \Theta$ and $\psi\equiv R\sin n_0 \Theta$. Then for the $n^{th}$ domain wall, the boundary conditions read
\begin{equation}
\phi(z=0)=0,
\end{equation}
\begin{equation}
\phi(z=\pm\infty)= \phi_0\cos\frac{(n-1)\pi}{n_0},\;\;\phi(z=\mp\infty)= \phi_0\cos\frac{n\pi}{n_0}, 
\end{equation}
and
\begin{equation}
\psi(z=\pm\infty)= \phi_0\sin\frac{(n-1)\pi}{n_0},\;\;\phi(z=\mp\infty)= \phi_0\sin\frac{n\pi}{n_0}, 
\end{equation}
where each line demonstrate two distinct possibilities for each sign. Obviously, the odd and even properties of the fields, which are crucial for the next steps, disappear. In order to restore the parity characteristics in boundary behavior one needs to rotate the fields by angel $\alpha=(n-\frac{n_0+1}{2})T$, where $n$ counts the number of the intended walls when we enumerate them counter clockwise from $\Theta=0$ and $T=\frac{\pi}{n_0}$ is the angular period of the potential. 
\begin{equation}
V=\frac{\lambda}{4}\left(R^2-\phi_0^2\right)^2+\frac{\epsilon}{2}R^2\sin^2 (\Theta^\prime=\Theta+\alpha).
\end{equation}   
Then, for even $n_0$, the potential remains the same since the rotation angel is an integer times the period. But for $n_0$ odd, further elaboration is needed. In the rotated cordinate the boundary conditions are 
\begin{equation}
\phi^\prime(z=0)=0,
\end{equation}
\begin{equation}
\phi^\prime(z=\pm\infty)= \pm\phi_0\sin\frac{\pi}{2n_0}, 
\end{equation}
\begin{equation}
\psi^\prime(z=\pm\infty)= \phi_0\cos\frac{\pi}{2n_0},
\end{equation}
which exhibit the desired parity. Therefore the next steps are eligible and the approximation becomes
\begin{equation}
\phi^\prime(z)=\phi_0\sin\frac{\pi}{2n_0}\tanh{\sqrt{\epsilon}z},\label{pdw}
\end{equation}
and
\begin{equation}
\psi^\prime(z)=\sqrt{\phi_0^2\sin\frac{\pi}{2n_0}-2\frac{\epsilon}{\lambda}}\sech{\sqrt{\epsilon}z}.\label{sdw}
\end{equation}
The validity bound of the above approximation becomes narrower by increasing $n_0$, since it requires $(\phi_0^2\sin\frac{\pi}{2n_0}-2\frac{\epsilon}{\lambda})>0$. One can return to the original coordinates by a simple reverse rotation, respectively;
\begin{equation}
\begin{bmatrix}
\phi\\
\psi
\end{bmatrix}
=
\begin{bmatrix}
\cos\left(\frac{n_0+1}{2}-n\right)\frac{\pi}{n_0} & -\sin\left(\frac{n_0+1}{2}-n\right)\frac{\pi}{n_0}\\
\sin\left(\frac{n_0+1}{2}-n\right)\frac{\pi}{n_0} & \cos\left(\frac{n_0+1}{2}-n\right)\frac{\pi}{n_0}
\end{bmatrix}
\begin{bmatrix}
\phi_0\sin\frac{\pi}{2n_0}\tanh{\sqrt{\epsilon}z}\\
\sqrt{\phi_0^2\sin\frac{\pi}{2n_0}-2\frac{\epsilon}{\lambda}}\sech{\sqrt{\epsilon}z}
\end{bmatrix}
\end{equation}
\section{Inflationary Analysis}
The proposed potential in an inflationary perspective should be
categorized within the "new inflation" models \cite{EPFL}, while the
inflation is expected to begin near the maximum where the fields
leave the origin. New inflationary scenarios received great
welcome because they do not have the common problems of the old
inflation \cite{old} in completion of inflation \cite{ppm,olive}.
In fact, old scenarios need bubble collisions but a new
inflationary scenario can end with a more realistic process of
oscillation around a minimum \cite{nil,nias,EPFL}. Here, we try to
scrutinize how (\ref{original}) works as an inflationary
potential, too.
\subsection{Inflationary Scenario 1 : An Analytical
Overview on Simple Symmetry Breaking Inflation} The extreme
situation occurs when epsilon is considered as a small perturbation parameter in the original symmetry breaking term and can be ignored for the
most of the process, then the potential is readily reduced to the
single potential the same as the simple symmetry breaking which
has been already explored in some respects \cite{olive}.
Henceforth, in order to have a measure for the remaining part of
our survey it is convenient to know the inflationary
characteristics of the potential when the asymmetric term is
ignorably small. To provide more similarity we can rewrite the
potential as
\begin{equation}
V(R)=\Lambda(1-\frac{R^2}{\phi_0^2})^2,\label{ssb}
\end{equation}
in which we have used
\begin{eqnarray}
R=(\phi^2+\psi^2)^{\frac{1}{2}},\\
\Lambda=\frac{{\lambda}{\phi_0}^4}{4}.
\end{eqnarray}
Obviously, this potential is now in the domain of "New Inflation",
in which the field rolls away from an unstable equilibrium, here
placed at the origin. Even after considering the whole potential,
this will remain the main theme of the analysis. We can proceed by
making a Taylor expansion keeping the leading terms and ignoring the
rest due to $R^4<\phi_0 ^4$ for the roll down path.
\begin{equation}
V(R)\approx\Lambda\left(1-(\frac{R}{\mu}) ^2\right), \label{app}
\end{equation}
where
\begin{equation}
\mu=\frac{\phi_0}{\sqrt{2}}.
\end{equation}
It is a well-known result that a potential in the form
$V(r)=\Lambda(1-(\frac{r}{\mu}) ^\alpha)$ works properly as an
inflationary potential for $\alpha=2$ if it is LFM (Large Field
Model) i.e. $r\geq M_{Pl}$ \cite{EPFL}. To be more precautious and
to have a measure for the coming procedure, let us see the case
more closely. The slow roll parameters \cite{cdm} are
\begin{equation}
\varepsilon=\frac{8M_{Pl} ^2 R^2}{(\phi_0 ^2 - R^2)^2}
\label{srp1}
\end{equation}
and
\begin{equation}
\eta=\frac{4M_{Pl} ^2 (3R^2-\phi_0 ^2)}{(\phi_0 ^2 - R^2)^2}.
\label{slp2}
\end{equation}
To estimate the field value at the end of inflation we require
$\epsilon_{end}\sim 1$ \cite{cdm}, then the appropriate solution
would be
\begin{equation}
R_{end}=\sqrt{4M_{Pl} ^2 +\phi_0 ^2 -2\sqrt{4M_{Pl} ^4 + 2M_{Pl}
^2\phi_0 ^2 }}. \label{rend1}
\end{equation}
Assuming $\alpha\equiv\frac{M_{Pl}}{\phi_0}< 1$ and utilizing the
Taylor expansion in favor of the leading terms we can write
(\ref{rend1}) as
\begin{equation}
R_{end}\simeq\phi_0\sqrt{1-2\sqrt{2}\alpha+4\alpha^2-2\sqrt{2}\alpha^3}.
\label{rstar}
\end{equation}
One can recognize that the end of inflation happens close to the
true minimum. Having an estimate for $R_{end}$, we are able to
obtain the e-folding interval between the time that cosmological
scales leave the horizon and the end of inflation \cite{cdm}
\begin{equation}
N\approx\int^{R_*}
_{R_{end}}{\frac{V}{V,_R}}dR=\left(\frac{1}{8}R^2-\frac{1}{4\alpha
^{2}} \ln R\right)^{R_*} _{R_{end}}=
-\frac{1}{4\alpha^2}\ln\left(\frac{R_*}{R_{end}}\right)+\frac{1}{8}(\frac{1}{R_{*}^2}-\frac{1}{R_{end}^2}),
\label{efold1}
\end{equation}
where we set $M_{Pl}=1$ for simplicity. Substituting (\ref{rstar})
in (\ref{efold1}) and making some straightforward approximation
again, one obtains
\begin{equation}
N\approx-\frac{1}{8\alpha^2}\left[(\frac{R_*}{R_{end}}-2)-(\frac{R_*}{R_{end}}-2)^2\right]+\frac{1}{8}(\frac{1}{R_{*}^2}-\frac{1}{R_{end}^2})
=\frac{1}{8\alpha^2}\left[(2-R_*\alpha)^2-2(1-\sqrt{2}\alpha)+R_*
^2 \alpha^2\right].
\end{equation}
One can solve the above equation with repsect to $r_*$ for the
only acceptable solution:
\begin{equation}
R_*=\frac{1-\sqrt{4N\alpha^2-\sqrt{2}\alpha}}{\alpha}.\label{tit}
\end{equation}
So far we supposed $\alpha<1$ to validate our approximation and
now $\alpha$ appears in the denominator. One can readily justify
that $R_*$ is a monotonically decreasing function of $\alpha$. It
implies that lowering the $\alpha$ raises the field value in which
the desired scale leaves the horizon. This statement is reasonable since smaller $\alpha$ leads to a decrease in the
slope of the potential with respect to the field $R$. This
provides us a straightforward method to find the maximum allowed
value for $\alpha$ in which $R_*$ coincides with the origin i.e.
$R_*=0$, then the answer will be
\begin{equation}
\alpha_{max}=\frac{\sqrt{2}+\sqrt{2+16N}}{8N}.
\label{apr}
\end{equation}
Therefore from (\ref{apr}), if one requires $N>50$, it roughly means
$\alpha<0.0743$. which is in accord with the previous
assumption about smallness of $\alpha$. To have an insight, we
have to emphasize that this upper bound for $\alpha$ coincides
with ignorable $R_*$ and undetectably low value for primordial
gravitational waves as it will become clear shortly.

\begin{figure}
\centering
\includegraphics[width=0.4\linewidth]{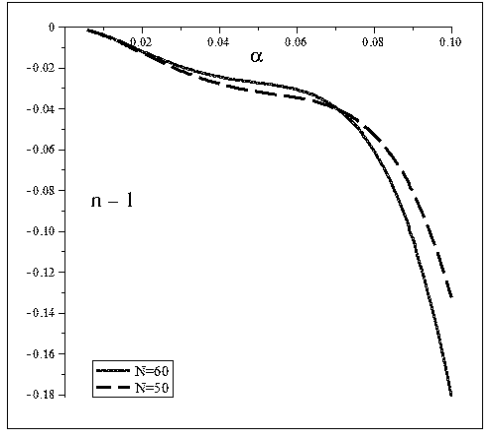}
\caption{This graph indicates $n_s-1$ with versus $\alpha$ for the
range of e-folding ($N_{min}=55$ and
$N_{max}=65$). Therefore one can claim that for the model under
investigation, spectral index is a monotonically decreasing
function of $\alpha$.} \label{fig:spectral}
\end{figure}

Now let us have a look at the most important observational
constraints on any inflationary hypothesis; spectral index and
tensor to scalar perturbation ratio. For the spectral index we
obtain
\begin{equation}
n_s -1\approx 2\eta_*
-6\varepsilon_*=-\frac{8(3R_*^2*\phi_0^2)}{(R_*^2-\phi_0^2)^2}\approx-{8\alpha^2}(1+5\alpha^2R_*^2)\approx-{8\alpha^2}\left[1+5(1-\sqrt{4N\alpha^2-\sqrt{2}\alpha}\,)^2\right].\label{bign}
\end{equation}
This estimation is accurate enough to indicate that for $\alpha=0$
one regains the scale-invariant Harrison-Zel'dovich-Peeble's
\cite{hzp} spectrum as expected. Recall that we keep assuming $M_{Pl}=1$. Let us use the fact that tensor to scalar perturbation
ratio "$r$"; $r=16\epsilon_*$ must be smaller than $0.11$
\cite{1}. Combining the definition of "$r$" with (\ref{srp1})
and making some simplification yields
\begin{equation}
r\approx\frac{128R_* ^2}{\phi_0^2(\phi_0^2-R_*^2)^2}.\label{ret}
\end{equation}
which points to a vanishingly small $r$ when the horizon exit
happens near the origin $(R_*=0)$ as mentioned before.
Substituting for $R_*$ we obtain
\begin{equation}
r\approx256\kappa^2\alpha^2(\kappa^2+\frac{1}{2}),
\end{equation}
where $\kappa$ is defined as
\begin{equation}
\kappa \equiv \pm\left[1-\sqrt{\alpha(4N\alpha-\sqrt{2})}\right].
\end{equation}
Then for the spectral index we have
\begin{equation}
n_s-1=-\frac{(1-9r_* ^4\alpha^4)}{16r_*
^2\alpha^4}r=\frac{(9\kappa^4-1)r}{16\kappa^2\alpha^2}
\end{equation}
This estimation is not precise enough yet but helps us to have a
better insight. Assuming a lower value for $\alpha$ means that
$R_*$ moves away from the origin but this can not effect the
gravitational wave strength considerably, since (\ref{ret}) can
always be approximated as
\begin{equation}
r\approx128 R_* ^2 \alpha^6. \label{ttosa}
\end{equation}
From (\ref{bign}) one can verify that $n_s-1$ is a monotonically
decreasing function of $\alpha$ in the allowed range
$\alpha<1/13.5$ (Figure  \ref{fig:spectral}).
But we have an accurate observation for the scalar spectral index
by Planck TT+lowP $68\%$ CL,  \cite{1,prcp};
\begin{equation}
0.9593<n_s<0.9717.
\end{equation}
This yields the range of valid $\alpha$ for a typical e-folding
$N=55$
\begin{equation}
0.046<\alpha<0.071.
\end{equation}
This means $14M_{Pl}<\phi_0<21.8M_{Pl}$ which exhibits a
relatively stringent fine tuned need of the model. On the other
hand, the estimated lower limit for $\alpha$ can help us to
determine the maximum of $R_*\times\alpha$ since from (\ref{tit})
we have
\begin{equation}
R_*\alpha=1-\sqrt{4N\alpha^2-\sqrt{2}\alpha}, \label{Rca}
\end{equation}
which is again a decreasing function of $\alpha$ in the permitted
region $0.046<\alpha<0.071$ such that $R_*\times\alpha$ remains
between $0$ and $0.04$, approximately. One can estimate the tensor
to scalar ratio using (\ref{Rca}) in (\ref{ttosa}) to obtain
\begin{equation}
r\approx128\alpha^2\left(1-\sqrt{4N\alpha^2-\sqrt{2}\alpha}\right)^2.
\end{equation}
\begin{figure}
\centering
{\includegraphics[width=.4\linewidth]{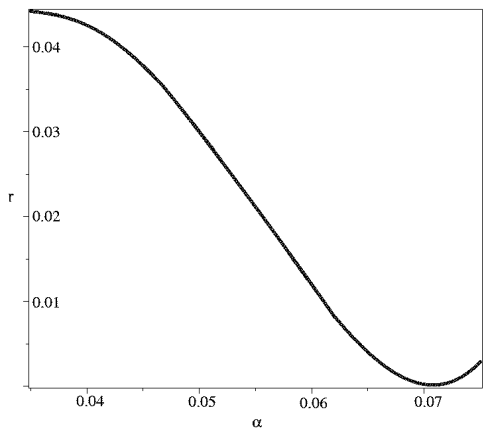}} \caption{Plotting
the approximate result for the tensor to scalar ratio shows that
for the target range of $\alpha$ one can claim $0<r<0.04$, which
is a typical behavior for a hill-top model.}
\label{fig:tensortoscalar}
\end{figure}
This function monotonically decreases with repsect to $\alpha$ in
the acceptable range of $\alpha$, with the maximum value about
$r=0.04$ for $\alpha_{min}=0.046$. As one expects, the tensor to
scalar ratio vanishes for $\alpha_{max}=0.071$ since it requires
the initial point to be the fixed point (Figure
\ref{fig:tensortoscalar}). Therefore we learn that the model needs
a super-Planckian value for $\phi_0$ to work properly according to
the available data from Planck+WP+BAO \cite{1,plc}. There is
another piece of observational information yet to be addressed.
Actually the oldest and the most well-known part of it. From the
expansion of the power spectra for the curvature perturbations we
have \cite{plc}
\begin{equation}
\mathcal{P}_{\mathcal{R}}(k)= A_s\left(\frac{k}{k_*}
\right)^{n_s-1+\frac{1}{2}dn_s/d\ln k \ln(k/k*)+...}.
\end{equation}
Thus, the Planck constraint on $r$ implies an upper bound on the
inflation energy scale
\begin{equation}
max(V_*)=\frac{3\pi^2 A_{s*}}{2}r_*M_{Pl}^4,
\end{equation}
which is readily transformed to the more useful form
\begin{equation}
\frac{H_*}{M_{Pl}}<3.7\,\times\,10^{-5}.
\end{equation}
Taking into account the slow-roll paradigm once more, we obtain
\begin{equation}
\sqrt{\frac{V_*}{3}}<3.7\,\times\,10^{-5}.
\end{equation}
which equivalently means
\begin{equation}
\lambda<3\,\times\,10^{-13}.
\end{equation}

\subsection{Inflationary Scenario 2: The Proposed Potential and Possible Inflationary Slow-rolls}
For the potential (\ref{original}) the dynamics is limited between
two extrema. The first is similar to the previous case where the field starts rolling down from the vicinity of the
$\psi=0$ axis and remains close to it throughout its motion. This
is the most probable possibility due to the angular minimum of the
potential and this is more or less the only possibility if one
takes $\epsilon>\lambda\phi_0 ^2$, because as it was discussed, in
this situation, the origin becomes a saddle point and $\psi=0$
seems like an attractor path. But if we assume
$\epsilon<\lambda\phi_0 ^2$, then other paths are possible
depending on different initial conditions, reminding that the
nearer to the $\psi$ axis the less chance for the path to be
chosen, again because of the angular behavior of the potential.
But as a possibility, though a very weak one, we consider the
ultimate radial path from the origin to one of the side saddle
points at $\psi=\pm\sqrt{\phi_0 ^2-\epsilon/\lambda}$ and then a
curved orbit toward the true vacuum. Note that the second part of
the path could happen independently from the starting point. At
the point where these different trajectories meet, some transient
oscillations might happen, which are damped by the following inflationary period. Generally, these two extreme paths could
bring different expansions as we will discuss. For the moment, let
us concentrate on the path from the origin toward the
$\psi=\sqrt{\phi_0 ^2 - \epsilon/\lambda}$ saddle point. Remember
that although this is not an attractor and the chance of following this path is
low, still as we emphasized earlier, we consider it as a bound
for what could happen. As long as we restrict ourselves to move on
the $\phi =0$ axis, the effective potential can be simplified as
\begin{equation}
V(\phi=0,\psi)=\frac{\lambda}{4}(\psi^2-\phi_0
^2)^2+\frac{1}{2}\epsilon\psi^2,
\end{equation}
which can be reordered as
\begin{equation}
V(\psi)= \frac{\lambda}{4}\left(\psi^2-(\phi_0 ^2-
\frac{\epsilon}{\lambda})\right)^2+\frac{\epsilon}{2}\left(\phi_0
^2-\frac{\epsilon}{2\lambda}\right).
\end{equation}
It can be recognized now that the potential is similar to
(\ref{ssb}) shifted by a constant. So we can proceed in a similar
manner unless this time we set $\psi_{end}=\sqrt{\phi_0 ^2 -
\epsilon/\lambda}$ to obtain the amount of e-foldings in this part
of the slow-roll path;
\begin{equation}
N\approx\frac{1}{M_{Pl} ^2}\int^{\psi_*}
_{\psi_{end}}{\frac{V}{V,_\psi}}d\psi\approx\frac{1}{4\alpha ^{2}}
\ln(\frac{\psi_{end}}{\psi_*})-\frac{1}{8}\psi_{end}^2.
\label{efold2}
\end{equation}
Actually, the inflation couldn't reach the saddle point at
$\phi=0$ and $\psi=\sqrt{\phi_0 ^2- \frac{\epsilon}{\lambda}}$
since it is interrupted at $\psi^2\approx(\phi_0
^2-\epsilon/\lambda)+2M_{Pl}(3M_{Pl}-\sqrt{2}\phi_0)$ due to the
growth of the slow-roll parameter $\eta$. This is a transitive
situation and inflation starts quickly in a new path as we will
consider shortly. The whole e-foldings through $\phi=0$ trajectory
is
\begin{equation}
N\approx\left(\frac{1}{8}\psi^2-\frac{\lambda\phi_0^4\ln\psi}{4(\lambda\phi_0^2-\varepsilon)}
+\frac{\varepsilon\ln(\lambda\phi_0^2-\lambda\psi^2-\varepsilon)}{4(\lambda\phi_0^2-\varepsilon)}
-\frac{\varepsilon^2\ln(\lambda\phi_0^2-\lambda\psi^2-\varepsilon)}{8\lambda(\lambda\phi_0^2-\varepsilon)}\right)^{\psi_*}
_{\psi_{end}}.
\end{equation}
To simplify the above expression, we invoke two facts; first we
know that the pivot scale leaves the horizon soon near the origin
($\psi_0\approx 0$) and second, the inflation stops essentially
before $\psi=\sqrt{\phi_0^2-\lambda/\epsilon}$ so the last two
terms make no considerable contribution and we finally obtain
\begin{equation}
N\approx-\frac{\lambda\phi_0^4\ln\psi_*}{4(\lambda\phi_0^2-\varepsilon)}.
\end{equation}
It is seen that the denominator could intensify the e-foldings by
providing a semi-flat trajectory on the $\phi=0$ axis. But
remember that this path is not likely to happen due to
instability. Now we can write
\begin{equation}
\psi_*\approx
e^{-4N\left(\frac{\phi_0^2-\epsilon/\lambda}{\phi_0^4}\right)}.
\end{equation}
However the effect of varying $\epsilon/\lambda\phi_0^2$ only
slightly changes our picture about $\psi_*\approx 0$ since
\begin{equation}
\lambda\phi_0^2>\epsilon>0 \Longrightarrow 0.37\,M_{Pl}<\psi_*<
M_{Pl}.
\end{equation}
All other features along this path more or less resemble the
ordinary symmetry breaking case which was discussed
earlier and we will consider this again from a different view. But
inflation could happen along a completely different path; starting
from the saddle point and ending at the true vacuum. To have an
estimation about the selected path due to its low kinetical energy
we suppose that the fields remain in the radial minimum throughout
their trajectory. Therefore for obtaining the equation of the
estimated path first we find the radial minimum of the potential
in the polar form (\ref{polar})
\begin{equation}
\frac{\partial\,V}{\partial\,R}=\lambda R (R ^2-\phi_0 ^2)+
\epsilon R  \sin^2 \Theta=0.
\end{equation}
Ignoring the $R =0$ solution, which correspond to the maximum at the
origin, for a given angle $\Theta$ the radial minimum obeys the
following equation
\begin{equation}
R _{min}^2(\Theta)=\phi_0 ^2-\frac{\epsilon}{\lambda}\sin^2
\Theta.
\end{equation}
Returning to the original Cartesian form, we have
\begin{equation}
\phi^2+\psi^2=\phi_0^2-\frac{\epsilon}{\lambda}\frac{\psi^2}{\phi^2+\psi^2}.
\label{path}
\end{equation}
Setting $\epsilon=0$ we recover the $U(1)$ symmetry as expected.
The true vacua also satisfy the above equation. To simplify the
analysis, we will focus on the $\psi>0, \phi>0$ quarter and solve
(\ref{path}) for $\phi^2$ to obtain
\begin{equation}
\phi^2=\frac{\phi_0 ^2}{2}-\psi^2+\frac{\phi_0
^2}{2}\sqrt{1-\frac{4\epsilon\psi^2}{\lambda\phi_0 ^4}}
\end{equation}
since we assume the $\epsilon\ll\lambda\phi_0 ^2$ regime,
expanding the square root and keeping the most important terms we
deduce
\begin{equation}
\phi^2=\phi_0 ^2-\psi^2\left(1+\frac{\epsilon}{\lambda\phi_0
^2}\right),
\end{equation}
or equivalently
\begin{equation}
\psi=\sqrt{\frac{\phi_0 ^2-\phi^2}{1+\frac{\epsilon}{\lambda\phi_0
^2}}}.\label{apppsi}
\end{equation}
Finally the above approximation allows us to recast the potential
(\ref{original}) in the form of a single field potential
\begin{equation}
V(\phi,\psi=\sqrt{\frac{\phi_0
^2-\phi^2}{1+\frac{\epsilon}{\lambda\phi_0^2}}})=\frac{\lambda\epsilon\phi_0^2(\phi_0^2-\phi^2)(3\epsilon+2\lambda\phi_0^2-\epsilon\frac{\phi^2}
{\phi_0^2})}{(\lambda\phi_0^2+\epsilon)^2},
\end{equation}
while for the supposed slow-roll path $\phi<\phi_0$ we estimate
\begin{equation}
V(\phi)\approx
\frac{\lambda\epsilon\phi_0^4(3\epsilon+2\lambda\phi_0^2)(1-\phi^2/\phi_0^2)}{4(\lambda\phi_0^2+\epsilon)^2}.
\end{equation}
Then obviously for $\lambda\phi_0^2\gg\epsilon$ we obtain
\begin{equation}
V(\phi)\approx\frac{\epsilon\phi_0^4}{2}(1-\frac{\phi^2}{\phi_0^2}).
\end{equation}
As we expected, the inflation energy scale is determined by
$\epsilon$ and $\phi_0$ since the height of the barrier depends on
them. Under this condition, the potential completely fits to the
first new inflationary models called hill-top models \cite{HT}
with the general shape
\begin{equation}
V(\phi)=\Lambda^4\left(1-\frac{\phi^p}{\mu^p}+\dotsb\right) ,
\end{equation}
if we set $p=2$,
\begin{equation}
\Lambda^4=\frac{\epsilon(2\phi_0^2\lambda-\epsilon)}{4\lambda} ,
\end{equation}
and
\begin{equation}
\mu=\phi_0.
\end{equation}
Then the model predictions are
\begin{equation}
n_s-1\approx\frac{-4M_{Pl}^2}{\mu^2}+\frac{3r}{8},
\end{equation}
and
\begin{equation}
r\approx\frac{32\phi_* ^2M_{Pl}^2}{\mu^4},
\end{equation}
which are in agreement with Planck+WP+BAO joint $95\%$ CL contours
for LFM ($\mu\gtrsim 9 M_{Pl}$) i.e.
\begin{equation}
\phi_0\gtrsim 9M_{Pl}.
\end{equation}
Although $\phi_0$ is reduced compared to the original symmetry
breaking case thanks to providing a longer curved trajectory with a
smaller slope, the potential is still considered a super-Planckian
model without any known physical motivation. Particularly, Picci-Quinn mechanism which is supposed to happen at QCD scale falls far bellow the needed energy of inflation and even the reheating process in the above hypothesis likely produces undesirable domain walls to explain the invisible axions. 

\begin{figure}\hspace{0.4cm}
\centering
\subfigure[]{\includegraphics[width=0.35\columnwidth]{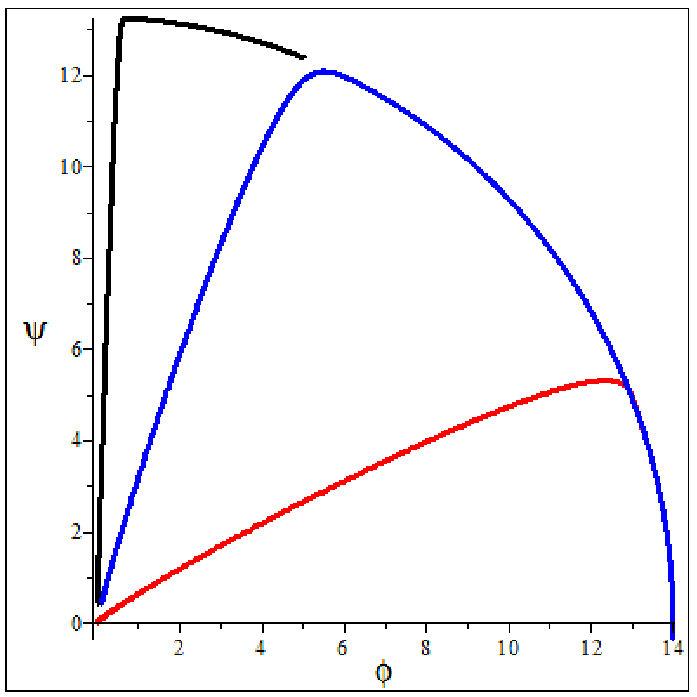}\label{fig:slowroll1}}
\subfigure[]{\includegraphics[width=0.35\columnwidth]{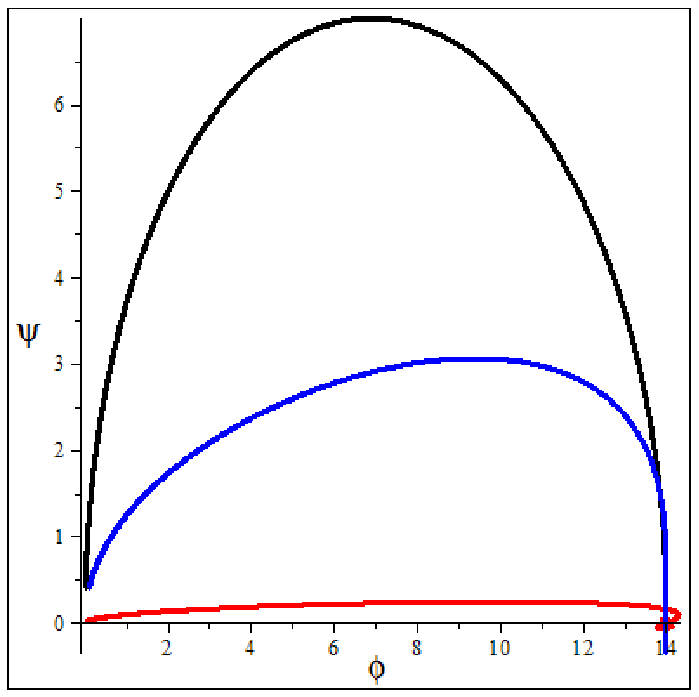}\label{fig:slowroll2}}
\caption{For $\phi_0\ll\frac{\epsilon}{\lambda}$, the slow-roll
path from the origin very well fits to a linear and a curved
distinct parts (left), but as this inequality becomes weaker, the path
exhibits more complexity (right), though still can be treated as the
previous case with a weaker approximation.} \label{fig:slowroll}
\end{figure}

The other probable scenario is rolling down the origin of field
space. First, let us see which direction is more likely to be
chosen for rolling down. Returning back to the field space polar coordinates
we obtain
\begin{equation}
\frac{\partial V(R,\Theta)}{\partial \Theta}=2\epsilon R^2
\sin\Theta\cos\Theta.\label{tet}
\end{equation}
For a certain $R$, the $\phi=0$ line is always maximum while
$\psi=0$ is a minimum so one probably expects slow-roll happen on
the $\phi$ axis or at least it appears as an effective attractor.
But as it is also obvious from (\ref{tet}), near the origin all
directions appear on the same footing so the field could follow
different trajectories later on. In other words, the path chosen
is very sensitive to the initial conditions and despite that the
$\phi$ axis is the most probable path, other options are not ruled
out. Simulations confirm this idea (Figure  \ref{fig:slowroll}). To
analyze the inflationary scenario in this case, first let us
derive $\frac{d \psi}{d
\phi}$ on the field space;
\begin{equation}
\frac{d \psi}{d
\phi}=\frac{\psi(\phi^2+\psi^2-\phi_0^2+\epsilon/\lambda)}{\phi(\phi^2+\psi^2-\phi_0^2)}
=\frac{\psi}{\phi}(1-\frac{\epsilon/\lambda}{\phi_0^2-\phi^2-\psi^2}).\label{dtod}
\end{equation}
Even if the ratio $\epsilon/\lambda\phi_0^2$ was not very small,
we would rely on the smallness of $\phi^2+\psi^2$ with respect to
$\phi_0^2$ to establish the following approximation
\begin{equation}
\frac{d \psi}{d
\phi}=\frac{\psi}{\phi}\left(1-\frac{\epsilon}{\lambda\phi_0^2}\right).
\end{equation}
The solution straightforwardly can be obtained as
\begin{equation}
\psi = A \phi ^ {1-\frac{\epsilon}{\lambda\phi_0^2}},
\end{equation}
in which $A$ is an arbitrary constant which stems from
arbitrariness in starting direction of the slow-roll. As
$\frac{\epsilon}{\lambda\phi_0^2}$ gets closer to 1, more
sophisticated approximations will be necessary. For example, the
above equation insures that the path outward origin remains linear
with a good accuracy for small $\frac{\epsilon}{\lambda\phi_0^2}$
ratio, until the path meets the radial minimum at
$\psi^2+\phi^2=\phi_0^2-\epsilon/\lambda$ in which $d\psi/d\phi$
quite suddenly vanishes. This abrupt redirection of course could
raise the chance of isocurvature density perturbations
\cite{Dod,AA} for large $\epsilon/\lambda\phi_0^2$ or even for small $\epsilon/\lambda\phi_0^2$ ratio at the end of inflation. But remember that expansion appears as a
damping term and prevents the field to acquire much kinetic energy
and the subsequent tumbling . In this regard, numerical
simulations confirm the slow-roll parameter predictions and
slow-roll continues up to the true vacuum vicinity despite the
slow-roll redirection in the field space (Figure
\ref{fig:slowroll1}). For now, let us see how the potential looks
in an arbitrary radial direction. To this end, we switch to the
polar coordinates once more and write the potential for a fixed
arbitrary angle $\Theta_0$.
\begin{equation}
V(R,\Theta=\Theta_0)=\frac{\lambda}{4}(R^2-\phi_0^2)^2+\frac{\epsilon}{2}R^2\sin^2{\Theta_0}=\frac{\lambda\phi_0^4}{4}\left(1-\frac{2R^2}{\phi_0^2}
(1-\frac{\epsilon}{\lambda\phi_0^2}\sin^2{\Theta_0})+\frac{R^4}{\phi_0^4}\right).
\end{equation}
Again, we encounter a hill-top model as could be expected and as
mentioned before we shouldn't worry about a slow-roll interrupt in
the redirection point since the inflation continues more or less
in the same manner up to the true vacuum. Note that in a typical
new inflation like what we consider, the first stages of inflation
are much more effective in producing the e-foldings and the
redirection always happens after these critical era such that
gives us enough excuse for the mentioned approximations. This time
we should define
\begin{equation}
\mu=\frac{\phi_0}{\sqrt{2\left(1-\frac{\epsilon}{\lambda\phi_0^2}\sin^2{\Theta_0}\right)}}.\label{mumu}
\end{equation}
To check this against the previous results, we can take
$\epsilon=0$ or $\Theta_0=0$ expecting that both of these choices
result in the constraint that we have already had from simple
symmetry breaking in (\ref{ssb}). Both of the above situations
reduce (\ref{mumu}) to
\begin{equation}
\mu=\frac{\phi_0}{\sqrt{2}}.
\end{equation}
Then the hill-top constraint of $\mu\gtrsim9$ readily yields
$\phi_0\gtrsim 12.72$ which is approximately in accord with the
previously achieved constraint. On the other hand, the bigger
$\epsilon\sin^2{\Theta}$, the smaller $\phi_0$ required for
matching with the observation. Here a delicate point has to be
taken into account. Although in the derivation of (\ref{mumu}) we
have not considered any constraint on the
$\frac{\epsilon}{\lambda\phi_0^2}$ ratio, but in order to have a
multi-directional operation, $\frac{\epsilon}{\lambda\phi_0^2}<1$
is required, since the other case has the ability to bring about
an imaginary $\mu$ in some directions which consequently changes
the sign of $\mu^2$ in (\ref{mumu}). Roughly speaking,
$\frac{\epsilon}{\lambda\phi_0^2}$ controls the range of the angle
$\omega$ in which the negative slope is seen from the origin: when
$\frac{\epsilon}{\lambda\phi_0^2}<1$ then $\omega=2\pi$ but in the
opposite case this angle for the positive half space $(\phi>0)$ is
obtained from
\begin{equation}
\frac{\omega}{2}=2\arcsin{\sqrt{\frac{\lambda}{\epsilon}}\phi_0}.
\end{equation}
The division by two here results from considering half plane. This
conclusion is not surprising at all since for
$\frac{\epsilon}{\lambda\phi_0^2}>1$ the origin becomes a saddle
point as discussed earlier. When the
$\frac{\epsilon}{\lambda\phi_0^2}>1$ inequality gets stronger, the
$\phi$ axis plays the role of attractor more effectively. So in
this regime we can ignore the $\psi$ field in the slow-roll stage
such that the potential reduces to
\begin{equation}
V(\phi,\psi=0)=\frac{\lambda}{4}(\phi^2-\phi_0^2)^2,
\end{equation}
which were discussed fully earlier. It might be worth arguing
that in the last case (i.e. $\frac{\epsilon}{\lambda\phi_0^2}>1$),
the potential for the initial condition in which fields are
located far from origin, imitates the chaotic inflation in the
form of $V(\psi)=\frac{\epsilon}{2}\psi^2$ which continues along
the trajectory to the true vacuum.
\begin{figure}\hspace{0.4cm}
\centering
\subfigure[]{\includegraphics[width=0.45\columnwidth]{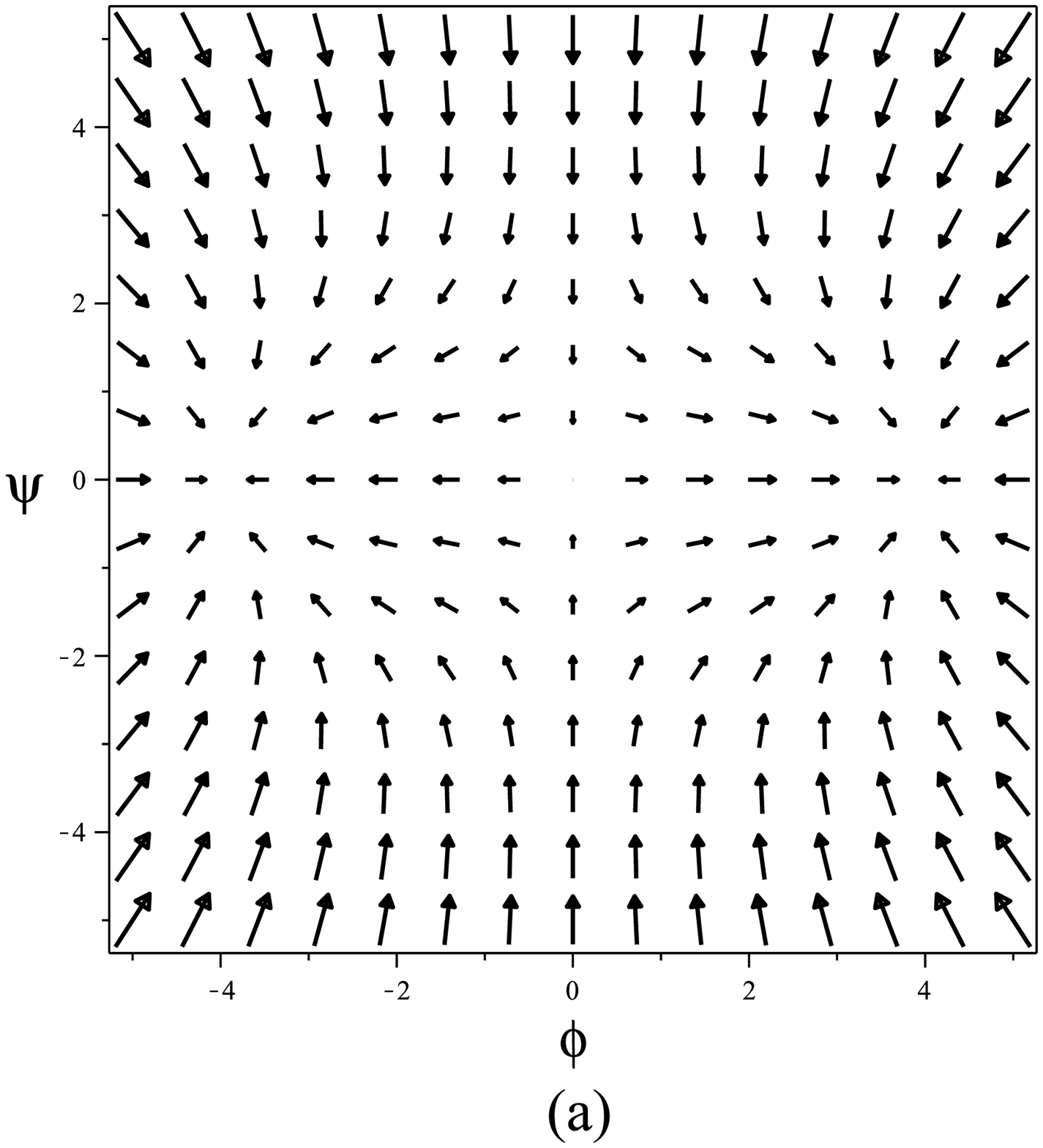}
\label{fig:sub1}}
\subfigure[]{\includegraphics[width=0.45\columnwidth]{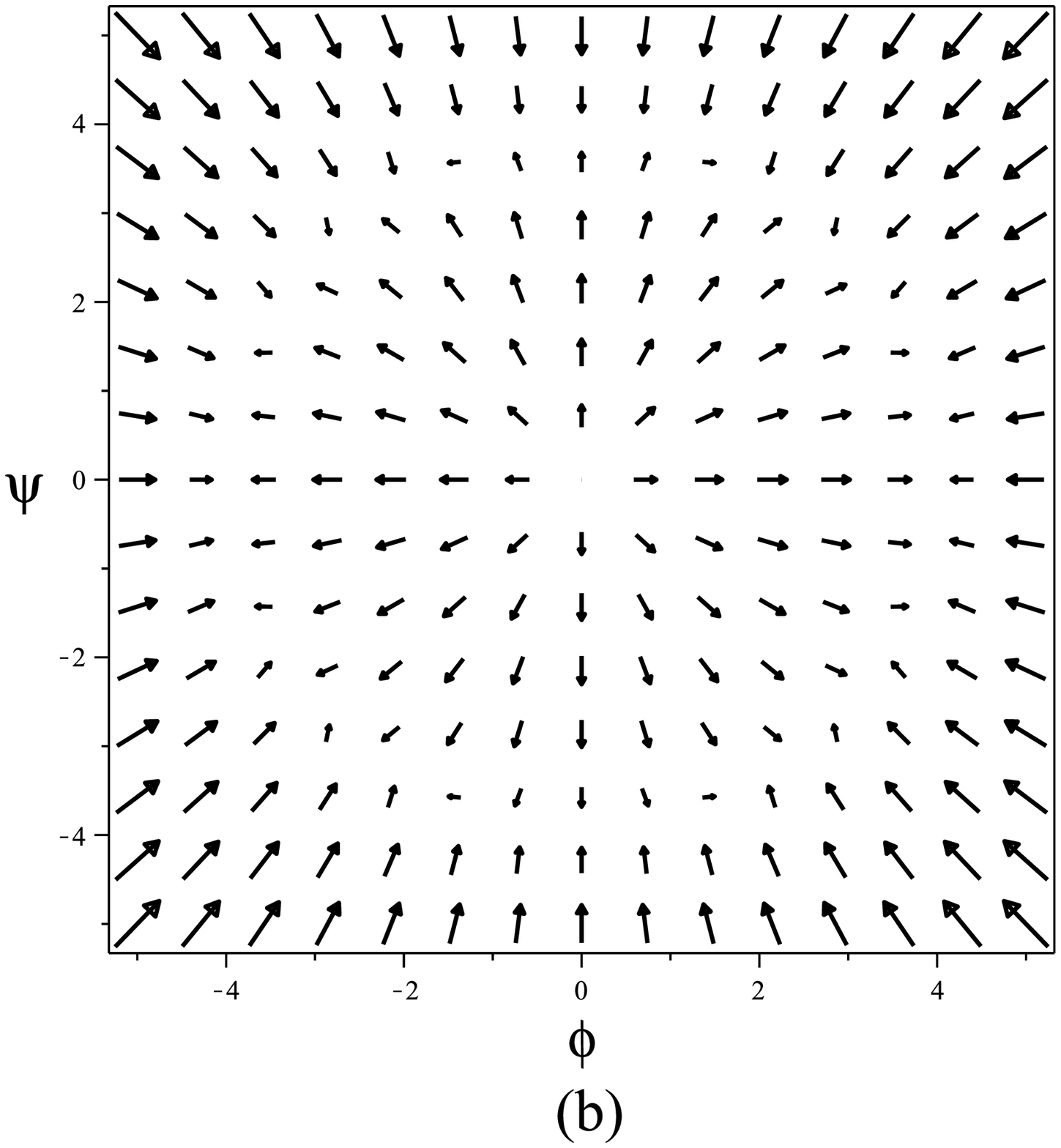} 
\label{fig:sub2}}
\caption{The above arrow diagrams show the direction of the gradient
vector field of the potential in the $\phi$-$\psi$ field
coordinates. In order to clarify the demonstration the
gradient vector field has been plotted in logarithmic scale. The
diagram on the left is plotted for $\lambda\phi_0 ^2<\epsilon$,
while the right one is for $\lambda\phi_0 ^2>\epsilon$. It is
obvious that the saddle points unite at the origin for the first
choice of the parameters.}
\label{mfnd}
\end{figure}

\begin{figure}\hspace{1.0cm}
\centering
\subfigure[]{\includegraphics[width=0.45\columnwidth]{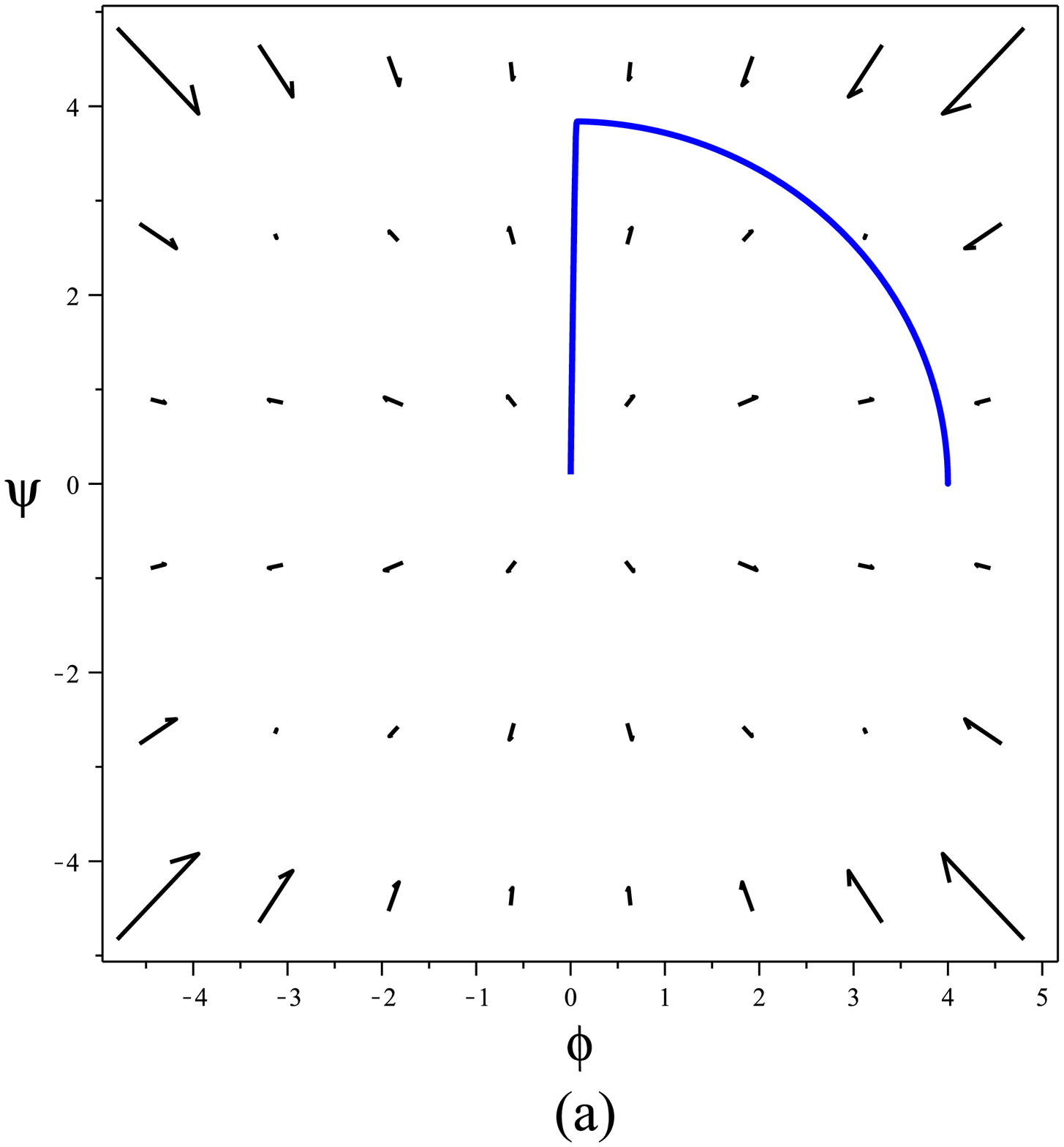}\label{fig:fieldflow1}}
\subfigure[]{\includegraphics[width=0.45\columnwidth]{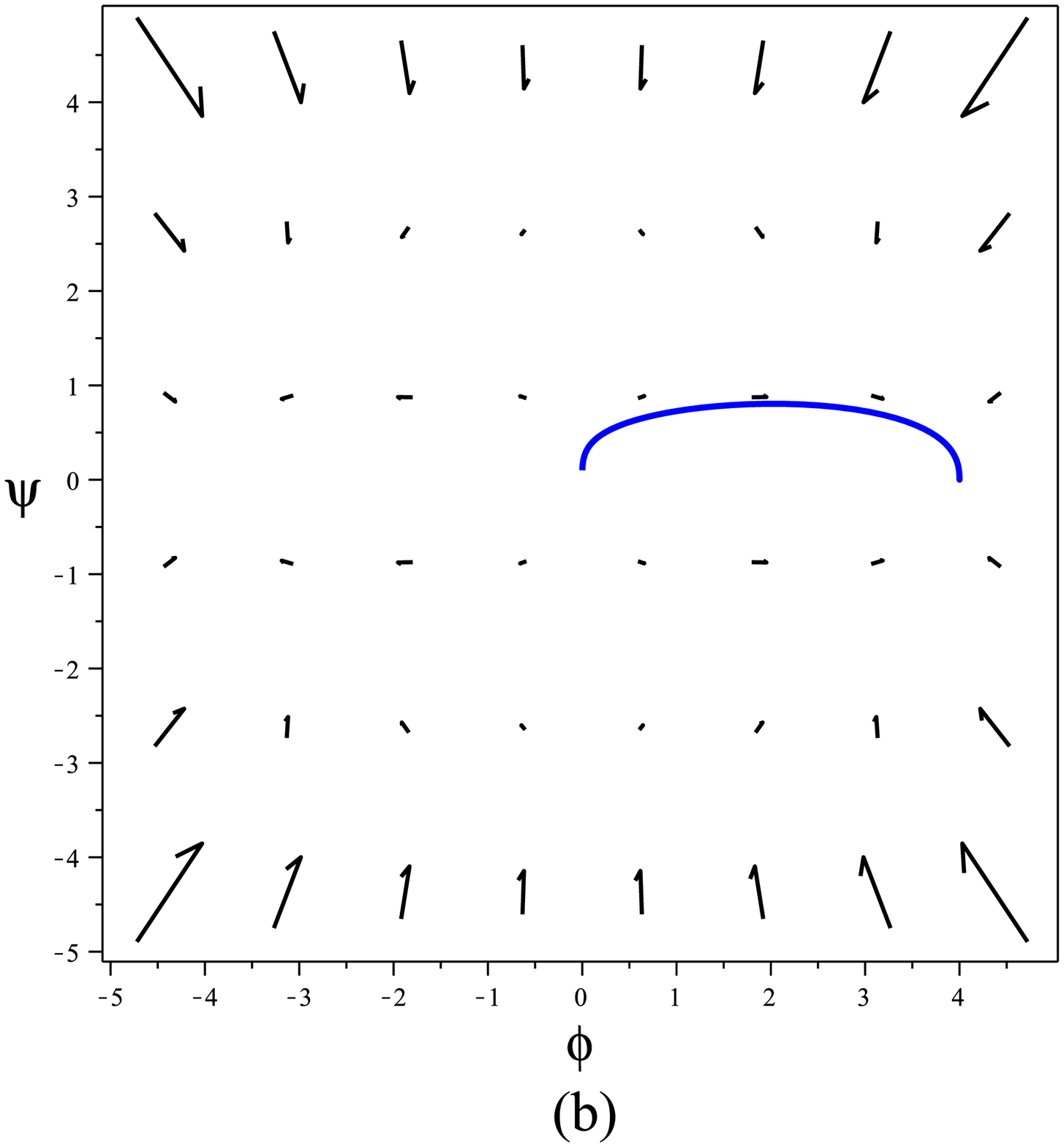}\label{fig:fieldflow2}}
\caption{The above graphs shows a sample line flow versus gradient
vectors. The left graph corresponds to the case
$\lambda\phi_0^2\gg\epsilon$ in which we have a central maximum and
two side saddle point. The right graph respectively correspond to
the case where $\lambda\phi_0^2\simeq\epsilon$ where the origin
turns to become a saddle point.}
\end{figure}

In Figure  \ref{fig:sub1} one observes the curved path from
the saddle point on the $\psi$ axis toward the true vacuum at
$(\phi=\phi_0,\psi=0)$. To plot these graphs we simply calculated
the absolute slope by
\begin{equation}
\vec{\nabla} V(\phi,\psi)=\frac{\partial V}{\partial
\phi}\hat{\phi}+\frac{\partial V}{\partial \psi}\hat{\psi}.
\end{equation}
To be more rigorous, the length of the
projected interval between the two saddle points on the
$\phi-\psi$ plane is $2\sqrt{\phi_0 ^2 - \epsilon/\lambda}$ and
disappears as soon as $\epsilon/\lambda$ reaches $\phi_0 ^2$ (or
gets bigger). Therefore, it is natural to take into account the
new path provided for the fields slow-roll. To indicate what we
exactly talk about one can refer to Figure \ref{fig:fieldflow1}.
As it is depicted in the figures, in the proposed potential, the
slow-roll path could be completely different from the radial flow
of an ordinary symmetry breaking. It is saying that the path is
very sensitive to the initial conditions and the figures are just
two possible paths among many.

Note that in Figure  \ref{fig:fieldflow1} the slow-roll comprises of two different paths; a radial and a nearly circular path, in contrast to the $\lambda\phi_0 ^2\ll\epsilon$ case, in which the saddle points unite at the origin. As
we have shortly discussed under the motivation topic, the latter case approximately reduces to the inverted hybrid inflation case \cite{MVHI}, with high energy scale domain walls which are formed before the inflationary procedure. Let us get back to the main trend and suppose $\lambda\phi_0 ^2
>\epsilon$. It was discussed earlier that the path consists of two distinct parts, a radial and a nearly
circular one, although the circular part changes in shape as the
inequality of $\lambda\phi_0 ^2
>\epsilon$ becomes weaker. One has to note that if the slow-roll
happens to be on the $\phi$ axis, then again the problem reduces
to the ordinary symmetry breaking scenario and the circular path
doesn't appear anymore. The remaining possibility that has less
importance relates to the case when $\phi_0^2$ is comparable to
$\epsilon/\lambda$. As it was discussed earlier, in this case the
slow-roll path is a curved line (Figure  \ref{fig:fieldflow2}) and
for such trajectories, there are other methods to deal with the
slow-roll process \cite{NG}-\cite{tran}. However, since recent
observations are generally in favor of single field inflation, or
at last, such models that evolve along an effectively single-field
attractor solution \cite{miap}, we limited our survey to the
situations in which we can approximate our double field potential
with a single field one.
\section{Conclusion}
We estimated the domain wall properties for an explicitly broken
$U(1)$ symmetric potential introducing an approximation that
nicely fits both the Euler-Lagrange equations with appropriate
boundary conditions and the static version of viral theorem. We
showed that adding one degree of freedom into our Lagrangian in
the form of a new field, helps us evade the domain wall
domination problem of the ordinary $\phi^4$ kink without decreasing the scale energy of the
potential. The price that has to be paid is
relaxing the $U(1)$ symmetry as an exact symmetry of the model. This allows us to have super-Planckian scale of energy for the peak of the potential while the domain wall energy is sufficiently low to
avoid conflict with observation. The exact allowed values of parameters are of course model dependent. More rigorously, it has to be first determined, in which cosmological era and correlation length (horizon), the domain walls will form. The descending of $U(1)$ to $Z_2$ is not an unprecedented scenario and the same explicit symmetry breaking have been suggested as a remedy for invisible axion and two Higgs doublet models. From an observational point of view, the model parameters could be set
such that the wall explains any confirmed CMB residual dipole anisotropy. In the other extreme, we proposed the domain wall production to happen before inflation and claimed that for super-Planckian values, this scenario could work properly. 

We thoroughly examined the potential as the source of inflation. We just focused on those cases which are reduced to a single field inflation since such models have received more appreciation after the Planck data. Our study indicated that all
the mentioned scenarios could be classified into the "new
inflationary models" and almost always into the hill-top subclass of it. We also introduced an analytic, though approximate proof
for the well-known simple symmetry breaking potential which
indicates nearly complete accordance with the previously obtained
numerical values. We tried to encompass all inflationary
possibilities of the potential. As briefly mentioned, there is
hope to explain the possible CMB dipole anisotropy by means of
domain walls, which could simultaneously solve the invisible axion problem. Therefore, it is important to make a compromise between
cosmological evidence, particle physics requirements and domain wall formation as we tried to do so.

\end{document}